\documentclass[fleqn,10pt]{wlscirep}
\usepackage[utf8]{inputenc}
\usepackage[T1]{fontenc}
\usepackage{multirow}
\usepackage{graphicx}
\usepackage{subcaption}
\usepackage{glossaries}
\usepackage{xurl}
\usepackage{hyperref}

\makeglossaries

\title{Merging synthetic and real embryo data for advanced AI predictions}

\author[1,+,*]{Oriana Presacan}
\author[2,+]{Alexandru Dorobanțiu}
\author[3]{Vajira Thambawita}
\author[8]{Michael A. Riegler}
\author[5]{Mette H. Stensen}
\author[5]{Mario Iliceto}
\author[6]{Alexandru C. Aldea}
\author[4,7,+]{Akriti Sharma}
\affil[1]{Faculty of Electronics, Telecommunications, and Information Technology, National University of Science and Technology Politehnica Bucharest, Bucharest, 061071, Romania}
\affil[2]{Department of Computer Science and Electrical Engineering, Lucian Blaga University of Sibiu, Sibiu, 550024, Romania}
\affil[3]{Department of Holistic Systems, SimulaMet, Stensberggata 27, 0170 Oslo, Norway}
\affil[4]{Department of Computer Science, Oslo Metropolitan University, Oslo, 0167, Norway}
\affil[5]{Volvat Spiren, Oslo, 0176, Norway}
\affil[6]{Faculty of Biotechnologies, University of Agronomic Sciences and Veterinary Medicine, Bucharest, 011464, Romania}
\affil[7]{Department of Validation Intelligence for Autonomous Software Systems, Simula Research Laboratory, Oslo, 0164, Norway}
\affil[8]{Cyber Security, Simula Research Laboratory, Oslo, 0164, Norway}
\affil[+]{these authors contributed equally to this work}
\affil[*]{orianapresacan@gmail.com}

\begin{abstract}
Accurate embryo morphology assessment is essential in assisted reproductive technology for selecting the most viable embryo. Artificial intelligence has the potential to enhance this process. However, the limited availability of embryo data presents challenges for training deep learning models. To address this, we trained two generative models using two datasets—one we created and made publicly available, and one existing public dataset—to generate synthetic embryo images at various cell stages, including 2-cell, 4-cell, 8-cell, morula, and blastocyst. These were combined with real images to train classification models for embryo cell stage prediction. Our results demonstrate that incorporating synthetic images alongside real data improved classification performance, with the model achieving $97\%$ accuracy compared to $94.5\%$ when trained solely on real data. This trend remained consistent when tested on an external Blastocyst dataset from a different clinic. Notably, even when trained exclusively on synthetic data and tested on real data, the model achieved a high accuracy of $92\%$. Furthermore, combining synthetic data from both generative models yielded better classification results than using data from a single generative model. Four embryologists evaluated the fidelity of the synthetic images through a Turing test, during which they annotated inaccuracies and offered feedback. The analysis showed the diffusion model outperformed the generative adversarial network, deceiving embryologists $66.6\%$ versus $25.3\%$ and achieving lower Fréchet inception distance scores.

\end{abstract}

\begin{document}
\flushbottom
\maketitle

\thispagestyle{empty}

\section*{Introduction}
Infertility affects millions as a disorder of the reproductive system in both men and women. According to the World Health Organization (WHO), approximately one in six people of reproductive age worldwide experience infertility \cite{who2024infertility}.As part of treatment, Assisted Reproductive Technology (ART) procedures involve inseminating multiple eggs retrieved from the ovary and then selecting the most suitable embryo for implantation in the uterus. Although it is the leading solution for infertility, ART has an average success rate less than $30\%$ \cite{ESHRE2023}. The success of ART depends directly on the quality of the embryo selected for transfer. However, visual evaluations by embryologists are subjective and prone to human error error \cite{Sundvall2013, Adolfsson2018}. Typically, ART data consists of images and videos. Artificial intelligence (AI) methods are commonly applied to tasks such as image and video classification and segmentation. Incorporating AI techniques can help embryologists add objectivity to the decision-making process and improve outcomes \cite{Riegler2021}. Furthermore, the number of ART treatments in Europe is increasing \cite{Gliozheni2022}, making the manual analysis of the growing volume of data increasingly time-consuming and resource-intensive \cite{Sciorio2022}.

AI-based image and video analysis techniques have been applied in both research and industrial settings to perform embryo evaluations and annotations on time-lapse data \cite{Thirumalaraju2021,Danardono2022,Dimitriadis2019,Liu2019,Lucio2022,TheilgaardLassen2023}. Indeed, the use of AI is gaining momentum as a promising technology to improve the efficiency of ART procedures \cite{Jiang2023}. According to the existing literature, convolutional neural network (CNN) and long short-term memory (LSTM) models are the most used AI approaches in ART. The application of CNNs for embryo assessments has been shown to outperform traditional methods \cite{Bormann2020}. However, advancements in AI now extend beyond CNNs and LSTMs \cite{AITransformers}. To fully harness the advancements of AI in ART, embryo data is essential, with both its quantity (volume) and quality (variety and veracity) serving as key factors. 

Embryo data encompasses images that capture the development of embryos at different cell stages and specific time points, highlighting the morphological changes throughout their developmental phases. The introduction of time-lapse imaging in ART laboratories has enabled comprehensive evaluation of embryo morphology by providing detailed insights into embryo development \cite{Cruz2011}. Embryo development is characterized by cell divisions, with each division representing a distinct stage, ranging from the 2-cell stage to later phases that exhibit distinct morphological features such as the morula and blastocyst \cite{gilbert2000early}. A morula is a stage in embryo development, typically observed around day four, and is characterized by a compact mass of 16–32 cells that are tightly packed and undifferentiated into layers. The increased cell-to-cell adhesion in morula sets the foundation for subsequent development into the blastocyst stage. The blastocyst is defined by a fluid-filled cavity (the blastocoel), an inner cell mass that later develops into the embryo, and an outer trophectoderm contributing to the placenta. Morula cells secrete fluid to form a small cavity, while the outer cells flatten and form tight junctions that separate the embryo’s interior from its exterior. Membrane channels then increase the salt concentration for drawing in water osmotically to expand the cavity. Concurrently, ongoing cell division enlarges the embryo with thinning of the zona pellucida until the blastocyst hatches, forming the inner cell mass and an expanding fluid-filled cavity \cite{eshreatlas2016}. Embryologists assess embryo quality by examining the morphological characteristics and duration of each developmental stage, ultimately selecting an embryo for transfer at either the cleavage or blastocyst stage \cite{Martins2017}. A cleavage-stage transfer involves transferring embryos on day two or three of development. Embryos on day two typically consist of two to six blastomeres, while on day three, they generally contain six to eight or more blastomeres. Additionally, embryologists consider the synchrony of cell divisions, which refers to the sequence of simultaneous divisions of all cells in an embryo. Perfect synchrony results in the embryo progressing through the 2-cell, 4-cell, and 8-cell stages. Since synchronous development is regarded as a positive indicator of embryo quality \cite{Lemmen2008}, this study focused exclusively on the 2-cell, 4-cell, and 8-cell stages for cleavage-stage transfers and the morula and blastocyst stages for blastocyst-stage transfers.

The primary challenge limiting the availability of embryo data arises from privacy and ethical concerns surrounding data sharing. Nearly all existing studies in ART have not made their embryo data publicly available \cite{Urcelay2023}. Furthermore, research frameworks often rely on customized datasets, resulting in subjective interpretations of ground truth that are specific to each framework \cite{Salih2023}. Study \cite{Khosravi2019} introduced STORK, an AI-based method for automated blastocyst quality assessment; however, access to their dataset was limited. In contrast, in study \cite{kanakasabapathy2021adaptive}, the authors publicly released their dataset of blastocyst and non-blastocyst embryos, captured using various imaging systems, from high-quality clinical time-lapse systems to lower-quality 3D-printed smartphone-based systems. Using an adaptive neural network, they demonstrated that a model trained on one dataset maintained high accuracy when tested on images from different datasets. Additionally, study \cite{kromp2023annotated} advanced AI for embryo selection by releasing a public dataset of static blastocyst images, annotated with Gardner criteria and clinical parameters, for deep learning model training. While these studies focused on the blastocyst stage, another study \cite{Gomez2022} provided a fully annotated public dataset of time-lapse embryo development videos, covering 16 developmental phases. Table \ref{tab:datasets} summarizes these publicly available embryo datasets and their characteristics.

\begin{table}[h]
\centering
\begin{tabular}{p{6cm}p{2cm}p{8.5cm}}
\toprule
 \textbf{Title}  & \textbf{Size} & \textbf{Description} \\ \midrule
Adaptive adversarial neural networks for the analysis of lossy and domain-shifted datasets of medical images \cite{kanakasabapathy2021adaptive} & 3,063 images & Annotated embryo images, classified into blastocyst and non-blastocyst categories, with quality levels labeled on a scale from 4 to 1.\\ \midrule
A time-lapse embryo dataset for morphokinetic parameter prediction \cite{Gomez2022}  & 704 videos   & Annotated embryo images capturing 16 key developmental events, from polar body appearance to blastocyst hatching, with frames labeled by their post-fertilization timing. \\ \midrule
An annotated human blastocyst dataset to benchmark deep learning architectures for in vitro fertilization \cite{kromp2023annotated} & 2,344 images & Annotated blastocyst images with expansion grade, inner cell mass, trophectoderm quality, expert agreement scores, and clinical data including age, oocyte metrics, pregnancy outcomes, etc. \\ \midrule
Ours & 5,500 images & Annotated embryo images depicting the 2-cell, 4-cell, 8-cell, morula, and blastocyst developmental stages, supplemented by synthetic images generated using advanced generative models.  \\ \bottomrule
\end{tabular}
\caption{Overview of other existing public annotated embryo datasets.}
\label{tab:datasets}
\end{table}

While publicly available datasets are important for benchmarking and reproducibility in ART, their limited size can restrict the development of robust deep learning models. Various approaches, such as federated learning \cite{Sheller2020} and synthetic data generation \cite{Murtaza2023}, have been developed to address the challenges posed by limited data volume. Indeed, the use of synthetic data in the medical domain is not new; it has successfully provided solutions to overcome the legal, privacy, and security barriers that restrict access to medical data. Deep convolutional generative addversarial networks (GANs) \cite{Radford2016} have been used to create synthetic images for skin lesion classification \cite{Behara2023} and to enhance the training of CNNs for improved pneumonia detection in chest X-ray images\cite{Porkodi2023}. State-of-the-art generative techniques such as the style-based GAN (StyleGAN)\cite{Karras2019} and the latent diffusion model (LDM)\cite{LDM_paper} are extensively explored for synthetic data generation in the medical field. Research has shown that StyleGAN can generate synthetic brain tumor data \cite{Abdalaziz2024} and 3D brain MRI images for enhancing diagnostic tools and medical research\cite{hong2021}. Additionally, LDMs effectively generated synthetic brain images with conditioning on various brain attributes\cite{Walter2022} and augmented biomedical datasets for microscopy \cite{Ivanovs2023}. 

We identified a few research studies that utilized generative AI to create synthetic images of human embryos. The first study, Human Embryo Image Generator Based on Generative Adversarial Networks (HEMIGEN) \cite{Dirvanauskas2019}, used a GAN to generate synthetic images of human embryos at the 1-cell, 2-cell, and 4-cell developmental stages. The authors included a qualitative assessment in which embryologists were asked to identify the number of cells in the synthetic images. However, it did not evaluate the visual quality of the generated images. Additionally, neither the code nor the dataset used in the study was made publicly available, limiting the reproducibility and further exploration of their work. A more recent study \cite{Cao2024} employed the StyleGAN model to produce synthetic images of human embryos at the blastocyst stage, achieving a Fréchet inception distance (FID) score of $15.2$ by fine-tuning a pre-trained model. This study also included a Turing test, where the synthetic images fooled embryologists $44.3\%$ of the time. Another study \cite{Chae2024} employed the LDM to generate embryo images at both the blastocyst stage (day five) and the cleavage stage (day three). To evaluate the model, Turing tests compared real and synthetic images, resulting in accuracies of 0.57 for day three and 0.59 for day five embryos. For day three, the evaluation focused on cell number, evenness, and fragmentation percentage, while for day five, the inner cell mass, trophectoderm, and overall blastocyst morphology were assessed. The study relied solely on expert evaluation and did not use any quantitative assessment methods. Although these studies used synthetic datasets to address data scarcity, they covered limited embryo stages and did not assess whether the synthetic images actually improved classification performance, thereby lacking a complete end-to-end solution. Classifying the morphological characteristics of embryo cell stages is a fundamental task that AI algorithms must be trained for, especially when applied to complex tasks such as assessing embryo quality or predicting pregnancy. Several studies have focused on single-frame embryo morphology, either targeting specific features like blastocyst formation \cite{Liao2021} or identifying multiple morphological events from the 1-cell to 4-cell stages \cite{Raudonis2019}. Recently, a study introduced a deep learning model capable of detecting 11 key embryo morphokinetic events from videos instead of individual images, enhancing embryo stage classification across various time-lapse platforms \cite{Canat2024}.

To provide a comprehensive end-to-end solution, our current work addresses these gaps by generating synthetic images across a broader range of embryo cell stages, including the 2-cell, 4-cell, 8-cell, morula, and blastocyst stages. These images were then used to train multiple classification models, tackling the critical aspect of data scarcity. Recent deep learning studies have demonstrated that augmenting real datasets with synthetic images can improve classification performance \cite{azizi2023syntheticdatadiffusionmodels, zhou2023trainingairimproveimage, 10.1145/3603715}. To further enhance the diversity of the synthetic dataset, we incorporated data generated from two distinct models: a GAN and a diffusion model. Given that these models employ different generation processes, we hypothesized that their images would introduce unique and potentially complementary features. This increased diversity could help the classification models generalize better across various image features, leading to better predictions. This aligns with findings from \cite{azizi2023syntheticdatadiffusionmodels}, which demonstrated that combining synthetic data from different diffusion models led to improved classification performance, and \cite{zhou2023trainingairimproveimage}, which highlighted that merging synthetic data from diverse generation techniques enhanced model robustness and generalization. Furthermore, to evaluate the quality of the generated images, we conducted a qualitative analysis using a custom-developed web application. Embryologists utilized the platform to classify images as ’real’ or ’fake,’ provide comments, and highlight areas of low fidelity in the synthetic images. This approach aimed to validate that the generated images are indeed realistic, represent key morphological characteristics of embryo cell stages, and can be reliably used for training AI models, thereby aligning with the second key aspect of embryo data: quality.

Below, we summarize our main contributions as follows:
\begin{itemize}
    \item Published a new open access annotated real embryo dataset (\url{https://zenodo.org/records/14253170}) to facilitate and advance research efforts in the field of ART.
    \item Employed two types of generative models, specifically the LDM and the StyleGAN, to generate synthetic images of embryos.
    \item Conducted a comprehensive qualitative assessment to validate the fidelity of the generated images, with evaluations performed by expert embryologists to ensure data quality.
    \item Improved the accuracy of embryo stage classification by integrating synthetic images into the training process of deep learning models, achieving higher accuracy even on an external dataset.
     \item Published a synthetic embryo dataset (\url{https://huggingface.co/datasets/deepsynthbody/synembryo_latentdiffusion}, \url{https://huggingface.co/datasets/deepsynthbody/synembryo_stylegan}) and the corresponding pre-trained generative models (\url{https://huggingface.co/deepsynthbody/synembryo_ldm}, \url{https://huggingface.co/deepsynthbody/synembryo_stylegan}) to augment existing data and support deep learning applications, thereby addressing data scarcity in ART.
\end{itemize}

\section*{Materials and methods}
\subsection*{Time-lapse incubation}
Embryologists evaluate embryo quality by assessing specific morphological parameters related to cell division and embryo stages \cite{ESHRE2016}. In this context, the morphological development present in embryo images captured during incubation provides a reference for embryo assessment \cite{Montag2011}. Additionally, stable culture conditions particularly concerning pH and temperature must be maintained during the incubation \cite{Lundin2020}. The introduction of time-lapse incubator (TLI) systems in ART laboratories has ensured such an optimal environment where embryologists can assess the morphokinetics without removing embryos from the incubator or exposing them to unstable culture conditions \cite{Sciorio2019}. 

A TLI system consists of three components: a microscope, imaging software, and an incubator, which allow for continuous imaging and a non-invasive evaluation of the embryo's quality \cite{Louis2021}. The system illuminates an embryo with a light source and magnifies embryo cells using an inverted microscope, capturing images with a digital camera at regular intervals across different focal planes. The software then displays a timed sequence of embryo development images, allowing embryologists to annotate morphology. The incubator also includes an `EmbryoSlide', a dish with multiple wells for culturing individual embryos separately.

\subsection*{Data}
The images used for training and evaluating the generative models originated from two sources. The first source was Volvat Spiren, a fertility clinic in Oslo, Norway. These images were extracted from embryo video frames recorded using a TLI system known as the Embryoscope$^{TM}$ (Vitrolife, Sweden), equipped with an EmbryoSlide containing 12 wells. The TLI system features a camera positioned beneath a 635 nm LED light source, which passes through Hoffman's contrast modulation optics. For each embryo, the system captured 8-bit images every 7–20 min across 3 to 5 focal planes, based on the settings of the older EmbryoScope system and EmbryoViewer software. The frames were extracted and labeled according to the cell stage they depicted, as identified by embryologists’ annotations. These annotations provide timings in hours post-fertilization, indicating the temporal position of each cell stage present in the video frames. Based on these timings, we selected frames corresponding to the specific cell stages of morula and blastocyst. The initial dataset consisted of 4,172 time-lapse videos capturing embryo morphokinetics. For our study, we excluded embryos with fragmentation rates above $15\%$. Fragmentation rates were available for 2,254 videos, of which 1,644 had fragmentation rates below $15\%$. Consequently, we used these 1,644 videos and included only frames from the central plane. Although the embryo is not always centered in the image and may be partially occluded due to its position in the EmbryoSlide well, the images in the dataset still accurately represent the embryo’s developmental stage.

The second data source is the publicly available dataset `Human Embryo Time-Lapse Video' \cite{Gomez2022}, also generated using the Embryoscope$^{TM}$ system with similar imaging parameters: a 635 nm LED light source, Hoffman’s contrast optics, captured every 10-20 minutes across 7 focal planes.We selected frames corresponding to the central plane only. In addition to the images, the dataset included class labels associating each image with the embryo developmental stages (cell stages, morula, and blastocyst). Since there was no information on fragmentation rates, we visually inspected the frames and selected those with low fragmentation for the 8-cell stage, as fragmentation cannot be assessed at the morula and blastocyst stages. The images also had issues with brightness changes and white spots caused by problems during image acquisition \cite{Gomez2022}.

Our final dataset consists of 5,500 embryo images, representing five developmental stages, 2-cell, 4-cell, 8-cell, morula, and blastocyst, with 1,100 images per stage. Due to variations in the number of images from each data source for different stages, Table \ref{tab:data_dist} provides a detailed breakdown of the image distribution from each source. For training the generative models (StyleGAN and LDM), 1,000 images per stage were used. The remaining 100 images per stage were reserved solely for testing the downstream classification models (VGG, ResNet, and ViT) and were not used in any way during the training of the generative models. The train/test split was performed at the sequence level, selecting a single representative frame for each stage of embryo development to ensure that frames from the same sequence did not appear in both sets. To expand the dataset, we augmented the real images with 5,000 synthetic images per stage generated using two different generative models, enhancing the training data for classification.

\begin{table}[ht]
\centering
\begin{tabular}{ccc}
\toprule
\multicolumn{1}{c}{\multirow{2}{*}{\textbf{Embryo Cell Stage}}} & \multicolumn{2}{c}{\textbf{Number of Images from Datasets}} \\ \cmidrule{2-3} 
\multicolumn{1}{c}{} & \multicolumn{1}{l}{\textbf{Volvat Spiren, Oslo}} & \textbf{Human embryo time-lapse video} \\ \midrule
2-cells  & \multicolumn{1}{c}{1100} & 0       \\ \midrule
4-cells   & \multicolumn{1}{c}{1100} & 0      \\ \midrule
8-cells & \multicolumn{1}{c}{850}  & 250      \\ \midrule 
morula   & \multicolumn{1}{c}{660}   & 440    \\ \midrule
blastocyst  & \multicolumn{1}{c}{600}  & 500  \\ \bottomrule
\end{tabular}
\caption{Distribution of embryo images across different developmental stages and data sources.}
\label{tab:data_dist}
\end{table}

To ensure the robustness and generalizability of our results, we evaluated our models on a publicly available subset of an external Blastocyst dataset \cite{Khosravi2019}. This subset includes 98 labeled images, which were captured using the EmbryoScope time-lapse system (Vitrolife, Sweden). Images were acquired using a single red LED (635 nm) every 20 min and were annotated by embryologists with quality labels: good-quality, fair-quality, and poor-quality.

\subsection*{Ethical Considerations}
In this study, we used human embryo videos collected at the Volvat Spiren fertility clinic in Oslo between 2013 and 2019. The Regional Committee for Medical and Health Research Ethics–South East Norway (REK) approved the data collection and use of these videos for research purposes. All methods were carried out in accordance with relevant guidelines and regulations, including the Declaration of Helsinki and the General Data Protection Regulation (GDPR). All patients gave informed consent for the intended research use of their data. However, the data were fully anonymized. Specifically, all patient-identifying information was removed prior to data processing, ensuring compliance with REK regulations and GDPR.

\subsection*{System Workflow}
\begin{figure}[ht]
\centering
\includegraphics[width=1\linewidth]{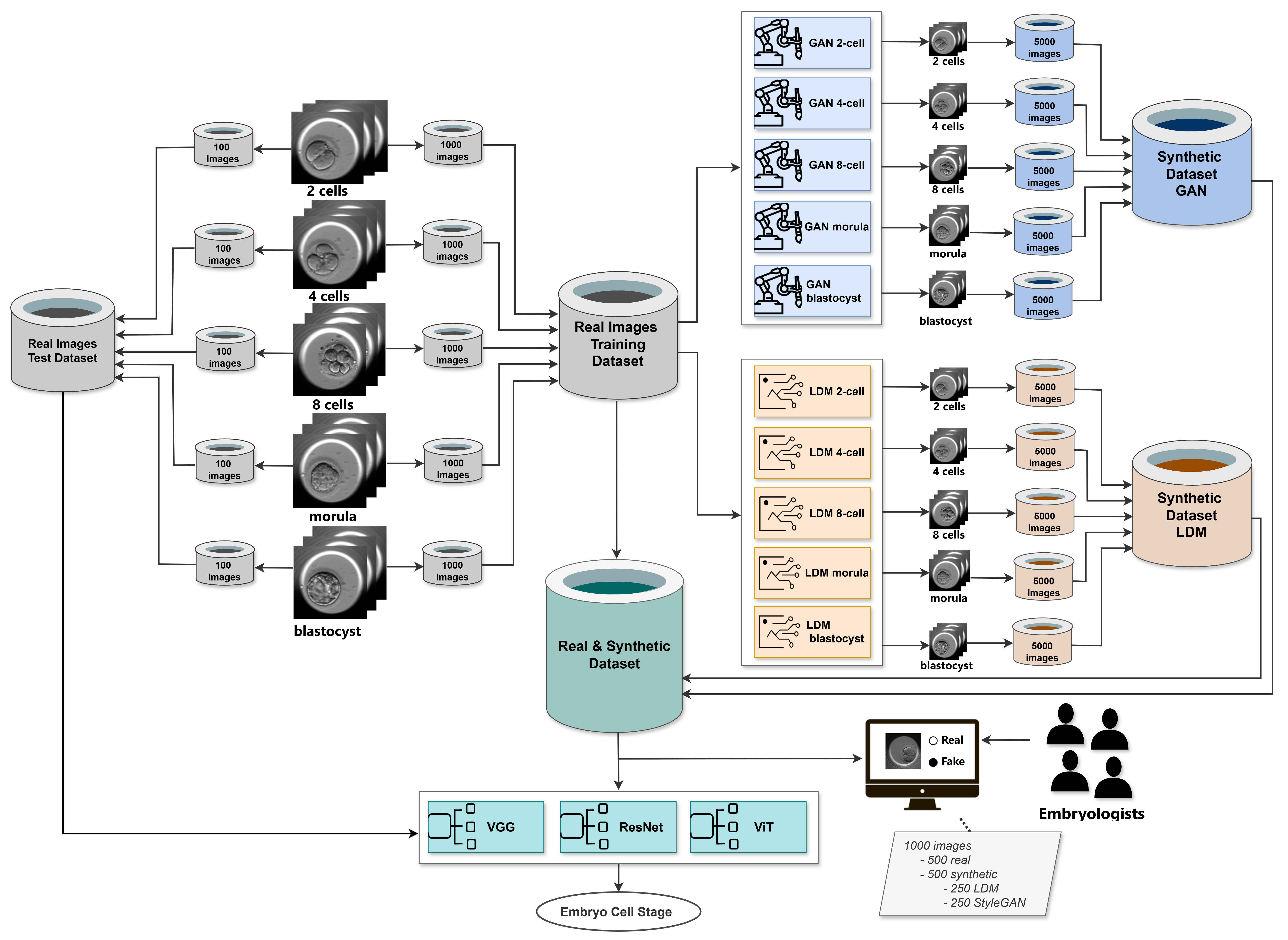}
\caption{Pipeline of our proposed system, encompassing the training of generative models, the generation of synthetic data, the training of classification models, and the qualitative assessment conducted by embryologists.}
\label{fig:pipeline}
\end{figure}

Figure \ref{fig:pipeline} illustrates the workflow of our system. For each embryo class, 2-cell, 4-cell, 8-cell, morula, blastocyst, 1,000 real images were used to train two types of generative models: GAN and diffusion model. A separate model was trained for each stage, resulting in a total of ten models. The best-performing model checkpoints, based on the lowest FID scores, were selected for image generation. From each selected model, 5,000 synthetic images were generated. These synthetic images, combined with the original real images, were then used to train three classification models: VGG, ResNet, and ViT. Model performance was evaluated using a separate set of 100 real images per stage and an external blastocyst dataset to ensure the generalizability and robustness of our results. Furthermore, four embryologists reviewed the generated images via a web application, identifying real versus synthetic samples and offering insights to confirm the visual quality of the data.

\subsection*{Generative Models}
Two generative models were selected for synthesizing images: a diffusion model, namely the LDM~\cite{LDM_paper} and a GAN, specifically StyleGAN~\cite{stylegan_paper}. These models were chosen not only for their well-established reputations in generative modeling but also for their distinct ways of generating images. The LDM transforms noise into high-quality images via a sequence of iterative steps, based on learned data distributions. On the other hand, the GAN uses an adversarial training methodology where two neural networks—the generator and discriminator—compete with each other: the generator creates images, while the discriminator evaluates them, driving the system towards producing precise results.

The LDM starts by reducing the dimensions of the input images using an autoencoder. This step improves computational efficiency while preserving the high quality of the images ~\cite{LDM_paper}. Subsequently, it proceeds to the diffusion phase, systematically introducing noise into the latent space. This converts the data into a noisy state through a predefined series of steps. A U-Net \cite{ronneberger2015unet} architecture is then used to perform the reverse process, which involves progressively removing the added noise at each step and transforming the noisy latent representation back into a clear image. During training, the model is exposed to examples of noisy data at various stages of the diffusion process, learning to predict earlier, less noisy states. Once trained, during the sampling phase, the model will be given pure noise from a normal distribution and will progressively transform this noise into a coherent image ~\cite{LDM_paper}.

StyleGAN operates based on the foundational principle of GAN, utilizing two networks: a generator and a discriminator. The generator's role is to produce realistic images given random noise. It starts this process by transforming the noise into a latent space representation via a mapping network. This representation is then adjusted to manipulate the generated image's style at various resolutions, employing a sequence of convolutional layers to enhance the image's resolution. The discriminator evaluates whether an image is real (from the dataset) or produced by the generator. Both the generator and discriminator undergo simultaneous training in a competitive manner, where the generator aims to fool the discriminator while the discriminator strives to become better at distinguishing real images from fakes ~\cite{stylegan_paper}.

The generative models were trained on 1,000 real images per class, each with an original resolution of 256×256 pixels. Training of the StyleGAN model was conducted using an NVIDIA RTX 4090, whereas the LDM utilized an NVIDIA Tesla V100 GPU. The LDM underwent unconditional training for 1000 epochs with a batch size of 16 and a learning rate of $0.000002$, as determined through repository code review and empirical testing. A linear noise scheduler was implemented for the diffusion process across 1000 steps. The pre-trained VQ-f4 autoencoder ~\cite{LDM_paper} was used to encode images into a 64x64x3 latent space. For StyleGAN, a pre-trained model based on the Flickr-Faces-HQ Dataset (FFHQ) \cite{ffhq-dataset} was employed after initial attempts at training from scratch yielded unsatisfactory results. Data augmentation techniques were also used to enhance the training process. The model underwent training for 1500 steps with a batch size of 16, a learning rate of $0.002$ for both the generator and discriminator, and an R1 regularization weight of 10. Detailed information about the configurations for both models used in this project is available in the project's GitHub repository, as well as links to the checkpoints of the models used to generate the final images.

To evaluate model performance throughout training, we calculated the FID every 50 epochs by generating 1,000 synthetic images and assessing their similarity to the 1,000 real training images. The training set was specifically used for FID calculation to track how well the model was learning the distribution of the training data. FID quantifies the difference in feature distributions between real and synthetic images, with lower scores indicating higher similarity. By monitoring FID scores, we can assess the authenticity of the synthetic images, ensuring they capture essential features from the real dataset. Ultimately, the model with the lowest FID score was selected for data generation. During sampling, random noise from a normal distribution was fed into the models, which converted them into images based on the learned patterns and features at various scales. A total of 5,000 synthetic images were generated for each class with each model and subsequently used for classification tasks and qualitative analysis.

\subsection*{Classification Models}

Three distinct models were employed for multi-class classification among five cell stages, including two CNNs—VGG and ResNet—and a transformer, ViT. These models were chosen for their established effectiveness in image classification tasks, with the goal of assessing whether a transformer-based architecture can outperform traditional CNN-based models. After experimenting with multiple model sizes, we identified VGG16, ResNet50, and ViT16 as the most effective architectures for our task. These models and their pre-trained versions were sourced from the \textit{torchvision} library, specifically \textit{vit\_b\_16}, \textit{vgg16\_bn}, \textit{resnet50}. The training approach for all three models was uniform, utilizing a batch size of 32, an Adam optimizer with an initial learning rate of 0.0001, a learning rate scheduler, and cross-entropy loss. Training was halted if there was no reduction in validation loss for 30 epochs, with the model showing the lowest validation loss being kept. All three models were trained on an NVIDIA RTX 4090.

For image processing, normalization was conducted using the mean and standard deviation of the dataset, and the images were resized to 224x224 pixels to accommodate the requirements of the ViT. We experimented with various combinations of images, including solely synthetic images produced by StyleGAN and the LDM in amounts ranging from $250$ to $5,000$, as well as combinations of real and synthetic images. We used $1,000$ real images to match the dataset size used to train the generative models, ensuring a consistent training process, and then progressively incorporated varying numbers of synthetic images to determine if this could enhance the outcomes. The models were then evaluated on a separate set of 100 real images reserved for testing. 

\section*{Results}
\subsection*{Data generation results}
\begin{figure}[!t]
  \centering
    \textbf{Real Images}\\
    \begin{subfigure}{0.18\columnwidth}
    \includegraphics[width=\textwidth]{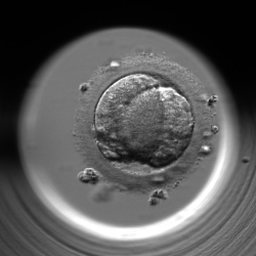} 
    \end{subfigure}
    \begin{subfigure}{0.18\columnwidth}
    \includegraphics[width=\textwidth]{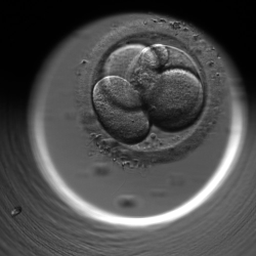} 
    \end{subfigure}
    \begin{subfigure}{0.18\columnwidth}
    \includegraphics[width=\textwidth]{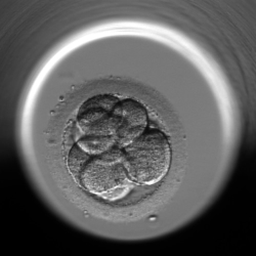} 
    \end{subfigure}
    \begin{subfigure}{0.18\columnwidth}
    \includegraphics[width=\textwidth]{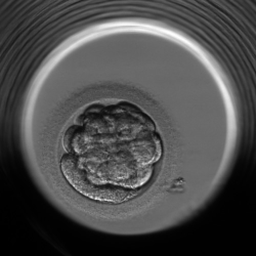} 
    \end{subfigure}
    \begin{subfigure}{0.18\columnwidth}
    \includegraphics[width=\textwidth]{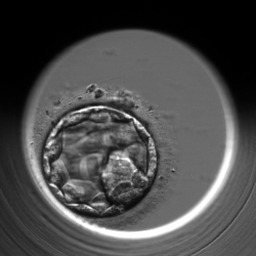} 
    \end{subfigure}\\ 
    \textbf{StyleGAN-generated Images}\\
    \begin{subfigure}{0.18\columnwidth}
    \includegraphics[width=\textwidth]{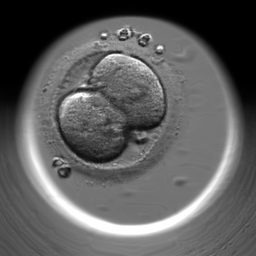} 
    \end{subfigure}
    \begin{subfigure}{0.18\columnwidth}
    \includegraphics[width=\textwidth]{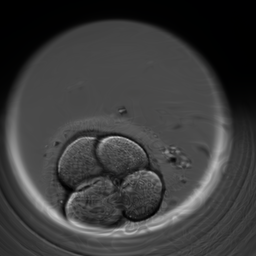} 
    \end{subfigure}
    \begin{subfigure}{0.18\columnwidth}
    \includegraphics[width=\textwidth]{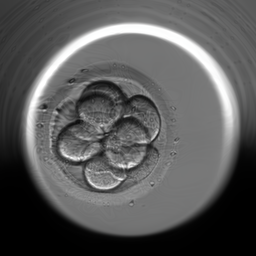} 
    \end{subfigure}
    \begin{subfigure}{0.18\columnwidth}
    \includegraphics[width=\textwidth]{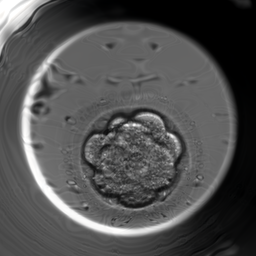} 
    \end{subfigure}
    \begin{subfigure}{0.18\columnwidth}
    \includegraphics[width=\textwidth]{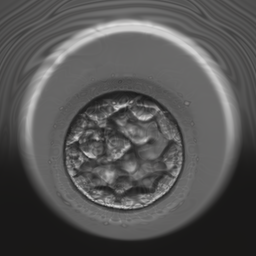} 
    \end{subfigure}\\ 
    \textbf{LDM-generated Images}\\
    \begin{subfigure}{0.18\columnwidth}
    \includegraphics[width=\textwidth]{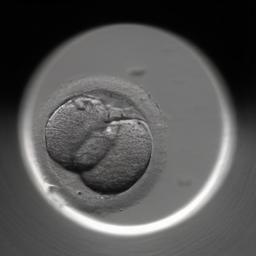} 
    \caption{2-cell}
    \end{subfigure}
    \begin{subfigure}{0.18\columnwidth}
    \includegraphics[width=\textwidth]{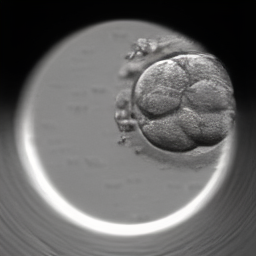} 
    \caption{4-cell}
    \end{subfigure}
    \begin{subfigure}{0.18\columnwidth}
    \includegraphics[width=\textwidth]{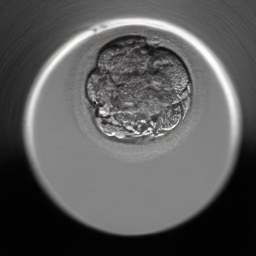} 
    \caption{8-cell}
    \end{subfigure}
    \begin{subfigure}{0.18\columnwidth}
    \includegraphics[width=\textwidth]{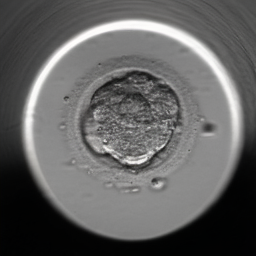} 
    \caption{morula}
    \end{subfigure}
    \begin{subfigure}{0.18\columnwidth}
    \includegraphics[width=\textwidth]{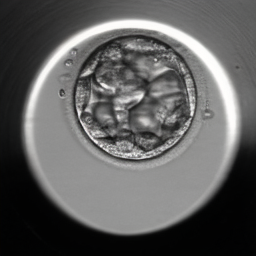} 
    \caption{blastocyst}
    \end{subfigure}
    \caption{Comparison of real versus synthetic images generated with StyleGAN and LDM models for each of the five embryo classes: 2-cell, 4-cell, 8-cell, morula, and blastocyst.}
  \label{fig:synthetic_img}
\end{figure}

Table \ref{tab:best_FID} presents the lowest FID scores achieved by StyleGAN and LDM models, which were trained for each embryo stage. The FID was calculated periodically throughout the training process, and the best checkpoints were selected based on the FID score. Overall, the LDM model demonstrates superior performance, as indicated by its lower FID scores. Figure \ref{fig:synthetic_img} illustrates real embryo images alongside synthetic images produced by StyleGAN and LDM for each of the five developmental cell stages: 2-cell, 4-cell, 8-cell, morula, and blastocyst.

\begin{table}[ht]
\centering
\begin{tabular}{cccccc}
\toprule
\textbf{Model} & \textbf{2-cell} & \textbf{4-cell} & \textbf{8-cell} & \textbf{blastocyst} & \textbf{morula} \\ \midrule
\textbf{LDM}     & 24    & 28    & 31    & 10    & 14     \\ \midrule
\textbf{StyleGAN} & 24    & 24    & 34    & 41    & 36     \\ \bottomrule
\end{tabular}
\caption{FID scores for the synthetic data generated by the best generative models trained for the different classes.}
\label{tab:best_FID}
\end{table}
\begin{table}[h!]
\centering
\begin{tabular}{ccccccccc}
\toprule
\multirow{2}{*}{\textbf{Model}} & \multirow{2}{*}{\textbf{Real}} & \multirow{2}{*}{\textbf{StyleGAN}} & \multirow{2}{*}{\textbf{LDM}} & \multicolumn{5}{c}{\textbf{Pre-trained}} \\ 
\cmidrule(l){5-9}
 & & & & \textbf{Accuracy} & \textbf{F1 Score} & \textbf{Precision} & \textbf{Recall} & \textbf{MCC} \\
\midrule
\multirow{2}{*}{\textbf{VGG}} & 0 & 5000 & 5000 & \textbf{93\%} & \textbf{0.93} & \textbf{0.93} & \textbf{0.95} & \textbf{0.92} \\
 & 1000 & 5000 & 5000 & \textbf{97\%} & \textbf{0.97} & \textbf{0.97} & \textbf{0.97} & \textbf{0.96} \\
\midrule
\multirow{2}{*}{\textbf{ResNet}} & 0 & 1000 & 1000 & 89\% & 0.89 & 0.90 & 0.89 & 0.87 \\
 & 1000 & 4000 & 4000 & \textbf{97\%} & \textbf{0.97} & \textbf{0.97} & \textbf{0.97} & \textbf{0.96} \\
\midrule
\multirow{2}{*}{\textbf{ViT}} & 0 & 2000 & 2000 & 91\% & 0.91 & 0.91 & 0.91 & 0.89 \\
 & 1000 & 4000 & 4000 & 93\% & 0.93 & 0.94 & 0.93 & 0.92 \\
\bottomrule
\end{tabular}
\caption{Best performance results for each of the three fine-tuned classification models—VGG, ResNet, and ViT—under two scenarios: training with only synthetic data and training with a combination of synthetic and real data. The table also shows the corresponding combinations of synthetic and real data used for training.}
\label{tab:classification_results}
\end{table}

\begin{figure}[htb!]
\centering
    \includegraphics[width=0.8\textwidth]{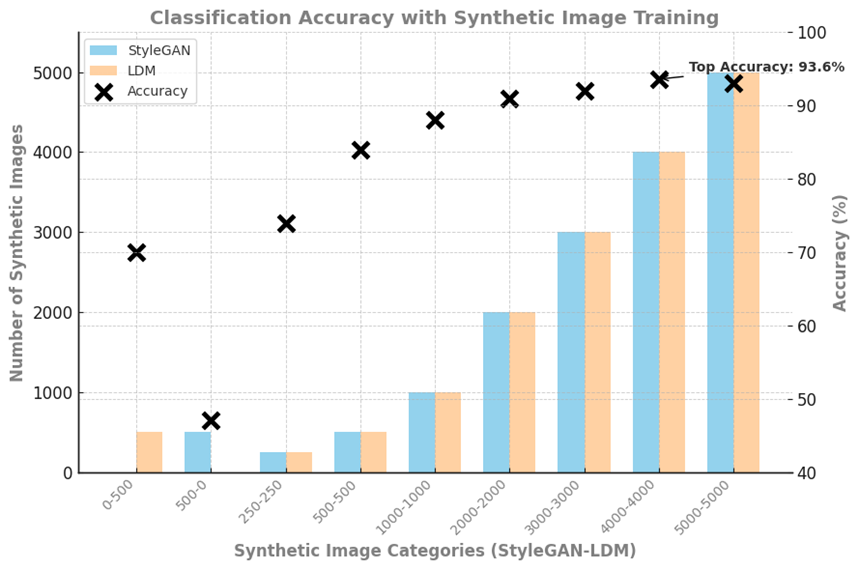} 
    \caption{Classification accuracy trends on test data (100 real images) for the VGG model, trained with various combinations of synthetic images generated by LDM and StyleGAN models.}
    \label{fig:acc_synthetic}
\end{figure}

\begin{figure}[htb!]
\centering
    \includegraphics[width=0.8\textwidth]{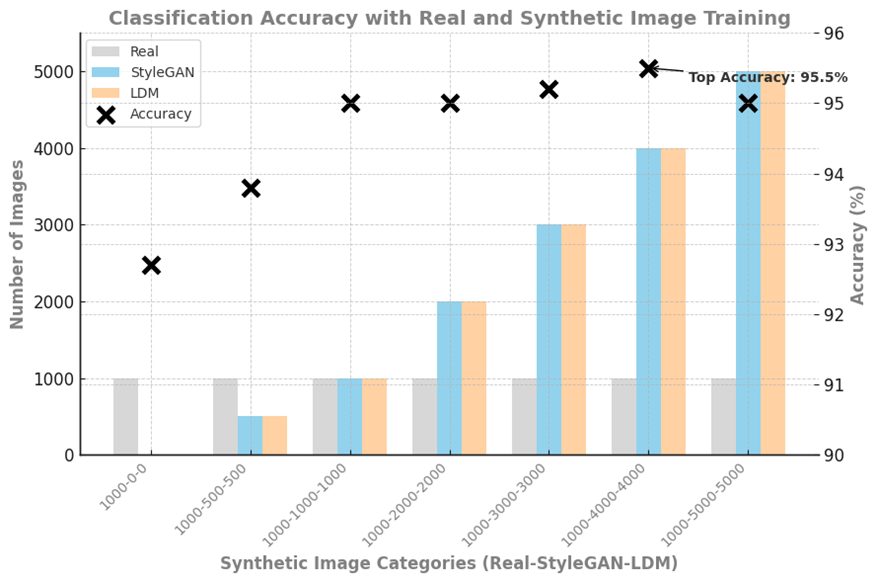 }
    \caption{Classification accuracy trends on test data (100 real images) trained with various combinations of real and synthetic data generated by LDM and StyleGAN models.}
    \label{fig:acc_synthetic_real}
\end{figure}

\begin{figure}[htb!]
\centering
    \includegraphics[width=0.7\textwidth]{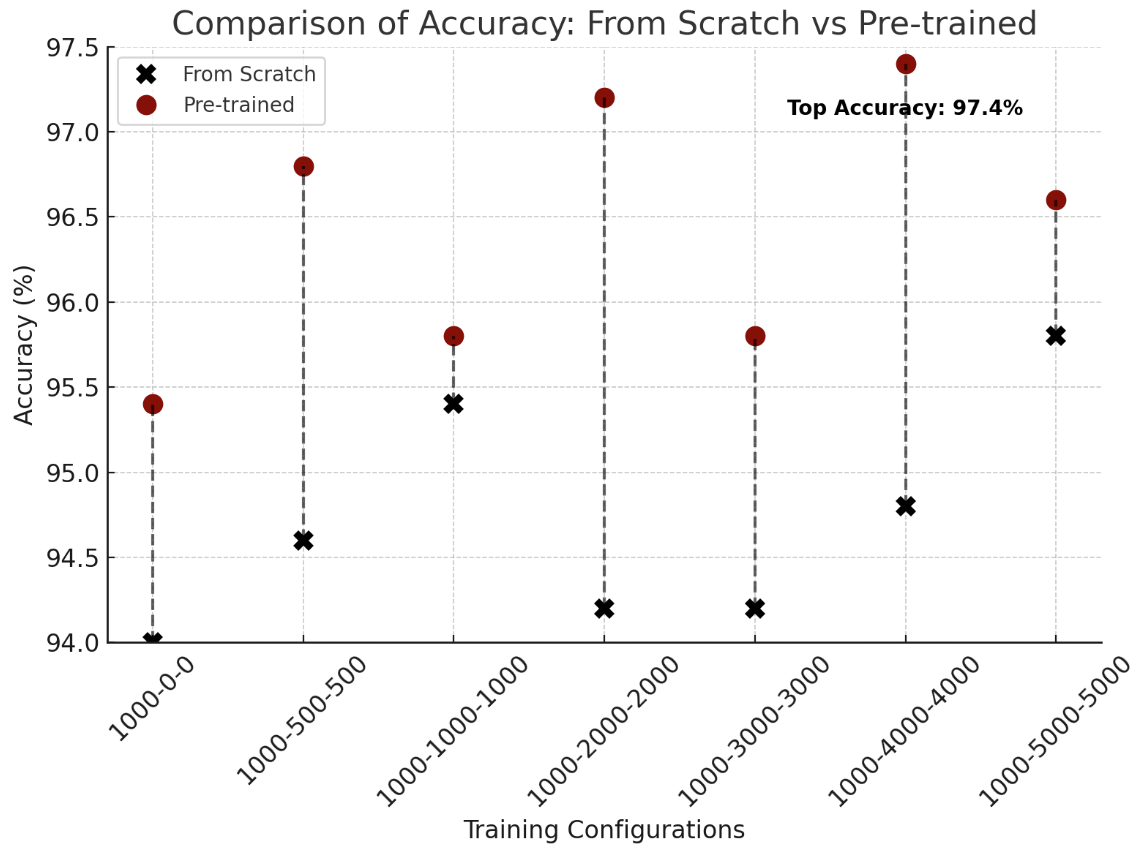}
    \caption{Accuracy differences between training the VGG model from scratch versus using a pre-trained model, based on the same data combinations as in Figure 4.}
    \label{fig:from scratch_pretrained}
\end{figure}


\subsection*{Classification results}
Table \ref{tab:classification_results} presents the classification results of the three models—VGG, ViT, and ResNet—trained on both real and synthetic data and evaluated on real data using various metrics, including accuracy, F1 score, precision, recall, and Matthews correlation coefficient (MCC). For clarity, we report only the best result for each model along with the corresponding combination of training data used, comparing two scenarios: training with synthetic data alone versus training with a combination of real and synthetic data. The VGG model demonstrated the best overall performance, and therefore, our subsequent analysis focuses exclusively on it. Comprehensive results for all three models are available in the "Supplementary Material".

To ensure the robustness and reliability of our results, we trained the VGG model using five different random seeds for initialization. We then computed the mean and standard deviation of the performance metrics to account for variability. Complete results for VGG models trained from scratch versus fine-tuned are in Tables S3 and S4 of the "Supplementary Material". Figure \ref{fig:acc_synthetic} illustrates the mean performance of the VGG16 model when trained solely on synthetic data. We employed different combinations of synthetic images generated by StyleGAN and LDM, varying these combinations to assess their impact on classification results. For results obtained from training with only LDM or only StyleGAN images, please refer to Tables S7 and S8 in the "Supplementary Material". Notably, training with 500 LDM images achieved an accuracy of $70\%$, compared to $47\%$ when using 500 StyleGAN images alone. This suggests that the LDM model might have generated higher-quality images. However, combining 250 StyleGAN images with 250 LDM images—maintaining a total of 500 images—increased the accuracy to $74\%$. This improvement indicates that integrating data from different generative models may enhance classification performance within the tested sample size range (up to 10,000 total synthetic images). While our dataset size limited testing at larger scales, this finding aligns with previous work on synthetic data combinations. Further testing with larger datasets are needed to confirm if this trend continues at greater scales but is out of scope of this work. The highest accuracy ($93.6\%$) was achieved by combining 4,000 StyleGAN images with 4,000 LDM images. Accuracy slightly decreased when using 5,000 GAN and 5,000 LDM images, possibly due to diminishing returns, where additional synthetic data introduces noise, subtle artifacts, or redundancy rather than meaningful variation. Figure \ref{fig:acc_synthetic_real} presents a similar results chart that includes training with 1,000 real images in addition to the synthetic ones. When training exclusively on real images, an accuracy of $92,7\%$ was obtained. In contrast, training with 1,000 real images combined with 4,000 LDM images and 4,000 StyleGAN images increased the accuracy to $95.5\%$. These results demonstrate that incorporating both synthetic and real data, especially from multiple generative models, enhances classification performance. Figure \ref{fig:from scratch_pretrained} shows the accuracy gain from training from scratch versus fine-tuning pre-trained models for the same data combinations from Figure \ref{fig:acc_synthetic_real}. Overall, the pre-trained VGG achieved the highest accuracy of $97\%$ when trained with 1,000 real images, plus 5000 synthetic images each from StyleGAN and LDM, compared to $94.5\%$ when trained solely on 1000 real images. This shows that synthetic images can effectively supplement real images to enhance model performance, which is valuable when real data is scarce or hard to obtain. However, the performance differences between different synthetic data combinations (e.g., 1000-2000-2000 and 1000-5000-5000) were not statistically significant, leaving the optimal amount of synthetic data an open question. Although the improvement is modest, it remains significant because even a slight enhancement can increase the model’s reliability and robustness, positively impacting user trust, qualities that are essential in the medical field.

Table \ref{tab:tab5} presents the mean accuracy, standard deviation, and confidence intervals for multiple model configurations evaluated on an external Blastocyst dataset. As expected, performance on the external dataset was lower than on our own dataset. However, we observed a substantial and statistically significant improvement when comparing the baseline configuration (1000-0-0) to those that included synthetic data. The best-performing synthetic data configuration (1000-4000-4000) achieved an accuracy of $84.69\%$ on the external dataset, marking a notable $15.81\%$ point increase over the baseline's $68.88\%$. This result highlights that our synthetic data approach not only enhances classification performance but also significantly improves model generalization to entirely unseen external datasets, an especially challenging task in medical image analysis. Moreover, all synthetic data configurations consistently outperformed the baseline, with the 1000-5000-5000 configuration demonstrating the most stable performance, as indicated by its narrower confidence interval (±2.84\%) compared to the baseline's (±5.86\%).

\renewcommand{\arraystretch}{1.4}
\begin{table}[ht]
\centering
\begin{tabular}{cccc}
\toprule
\textbf{Real} & \textbf{StyleGAN} & \textbf{LDM} & \textbf{Accuracy (95\% CI)} \\
\midrule
1000 & 0    & 0    & 68.88\% ± 5.86 (63.137, 74.623) \\ 
1000 & 500  & 500  & 77.55\% ± 6.687 (71.004, 84.096) \\
1000 & 1000 & 1000 & 74.98\% ± 6.949 (68.179, 81.781) \\
1000 & 2000 & 2000 & 73.47\% ± 6.432 (67.169, 79.771) \\
1000 & 3000 & 3000 & 76.27\% ± 7.622 (68.803, 83.737) \\
1000 & 4000 & 4000 & 84.69\% ± 6.828 (78.007, 91.373) \\
1000 & 5000 & 5000 & 78.08\% ± 2.844 (75.297, 80.863) \\
\bottomrule
\end{tabular}
\caption{Classification results (mean accuracy ± standard deviation) and confidence intervals (CI) for models trained on various real-synthetic data combinations and, evaluated on an external Blastocyst dataset.}
\label{tab:tab5}
\end{table}

\subsection*{Qualitative analysis}
We developed a web application to present the synthetic images to four human experts (embryologists), employing a Turing test framework. Each image was presented individually, and the embryologists had to determine whether it was real or fake. The goal was to evaluate the fidelity of the synthetic images and determine if they could mislead the experts. A total of 1,000 images were presented, comprising 500 real images (100 images per cell stage: 2-cell, 4-cell, 8-cell, morula, blastocyst) and 500 synthetic images. The synthetic images were evenly split between those generated by GAN (250 images) and LDM (250 images), with 50 images per cell stage. The application also allowed embryologists to mark areas of the images they considered unrealistic and included a text box for additional comments. A figure illustrating the application interface is provided in the "Supplementary Material".

\begin{figure}[htb!]
\centering
\includegraphics[width=\textwidth]{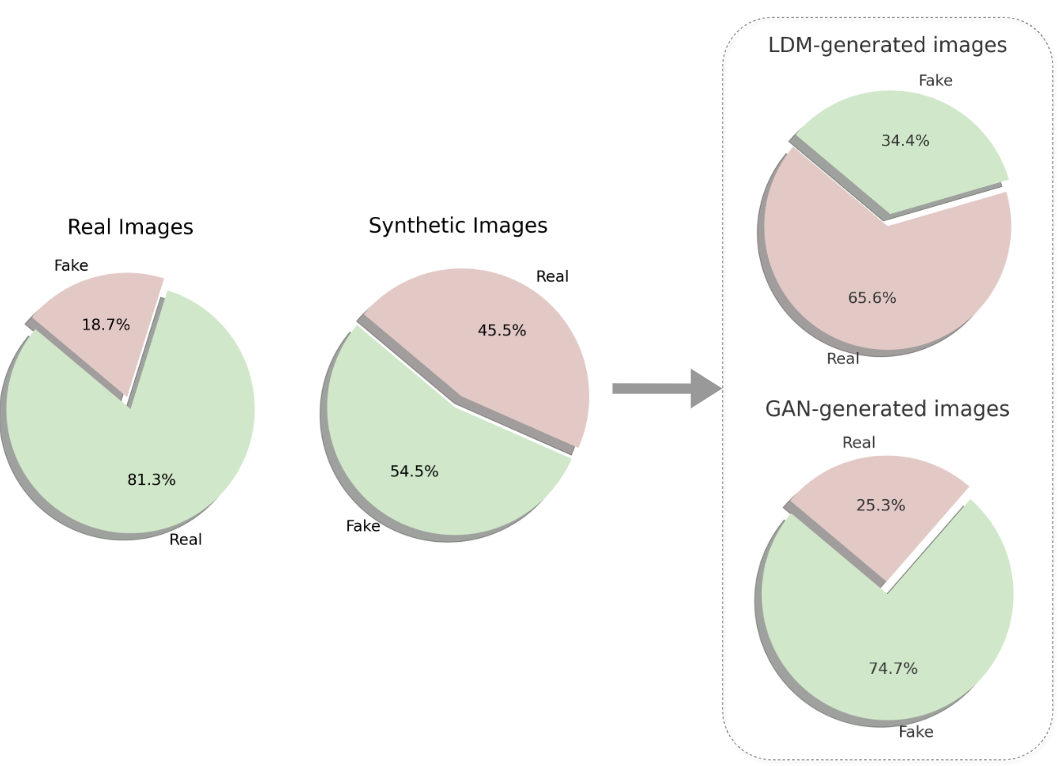}
\caption{Average Turing test results conducted by embryologists, evaluating their accuracy in identifying real and synthetic images. The "Real Images" pie chart displays the proportion of images correctly identified as real versus incorrectly labeled as fake. The "Synthetic Images" chart (covering both StyleGAN and LDM-generated images) illustrates the proportion of fake images accurately recognized as fake versus those mistakenly considered real. The final two charts break down these results by StyleGAN and LDM-generated images, respectively.}
\label{fig:pie}
\end{figure}

The averaged results of the Turing test conducted by the four embryologists are shown in Figure \ref{fig:pie}. We divided the outcomes into two categories: real images and generated images. The first chart shows the proportion of real images accurately identified as real, at $81.3\%$, versus those mistakenly classified as generated. This result is noteworthy, as the embryologists had initially anticipated a balanced, 50-50 distribution, citing that they often felt as though they were merely guessing. The second chart offers a similar comparison for synthetic images, revealing that the embryologists correctly identified only $54.5\%$ of these images. To assess the differences between the two generative models, we further divided the synthetic images into StyleGAN and LDM categories. This clearly shows that images generated by LDM appeared more realistic, as they were harder to classify correctly as "real" or "fake." Specifically, the embryologists correctly identified only $34.4\%$ of LDM-generated images, compared to $74.7\%$ of StyleGAN images. For a more detailed overview, Table \ref{tab:accuracy} shows the average accuracy with which the embryologists correctly identified real images as real and LDM-generated and  StyleGAN-generated images as fake, across the five cell stages. However, no single category stands out as consistently performing best. Overall, we can see that while embryologists are generally good at identifying real embryos, their accuracy in detecting synthetic images varies significantly depending on the model used to generate these images.

\begin{table}[ht]
\centering
\begin{tabular}{lccccc}
\toprule
\textbf{} & \textbf{2-cell} & \textbf{4-cell} & \textbf{8-cell} & \textbf{morula} & \textbf{blastocyst} \\ \midrule
\textbf{Real Images}  & 80\%  & 84\%  & 89\%   & 79\%    & 75\%     \\ \midrule
\textbf{StyleGAN Images}     & 76\%  & 77\%   & 73\%  & 71\%  & 86\%      \\ \midrule
\textbf{LDM Images}    & 37\%    & 40\%   & 29\%  & 38\% & 27\%     \\ \midrule
\end{tabular}
\caption{The average accuracy with which embryologists identified synthetic images as fake and real images as real, across the five cell classes.}
\label{tab:accuracy}
\end{table}

\begin{figure}[htb!]
  \centering
    \begin{subfigure}{0.25\columnwidth}
    \includegraphics[width=\textwidth]{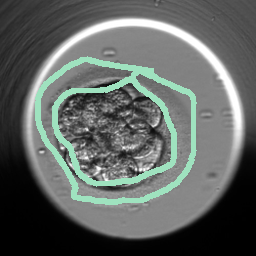} 
    \caption{Real}
    \label{fig:ann1}
    \end{subfigure}
    \begin{subfigure}{0.25\columnwidth}
    \includegraphics[width=\textwidth]{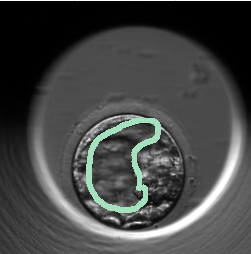} 
    \caption{Real}
    \label{fig:ann2}
    \end{subfigure}
    \begin{subfigure}{0.25\columnwidth}
    \includegraphics[width=\textwidth]{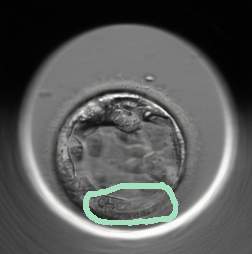} 
    \caption{LDM}
    \label{fig:ann3}
    \end{subfigure}
    \begin{subfigure}{0.25\columnwidth}
    \includegraphics[width=\textwidth]{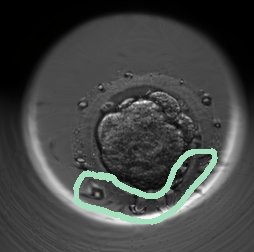} 
    \caption{GAN}
    \label{fig:ann4}
    \end{subfigure}
    \begin{subfigure}{0.25\columnwidth}
    \includegraphics[width=\textwidth]{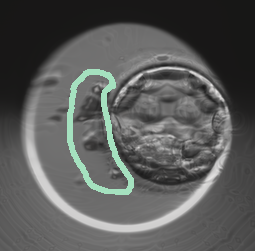} 
    \caption{GAN}
    \label{fig:ann5}
    \end{subfigure}
    \begin{subfigure}{0.25\columnwidth}
    \includegraphics[width=\textwidth]{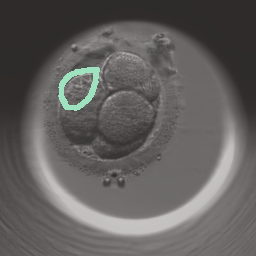} 
    \caption{GAN}
    \label{fig:ann6}
    \end{subfigure}
\caption{Annotations by embryologists highlighting perceived inaccuracies in images identified as synthetic. The first two images (a) and (b) depict real images, the following one (c) is generated using the LDM model, and the last three images (d), (e), and (f) are produced with the StyleGAN.}
  \label{fig:annotates_images}
\end{figure}

Figure \ref{fig:annotates_images} showcases examples of annotations made by embryologists on embryo images they perceived as synthetic, marking specific image regions or morphological features they identified as "fake." Several annotations focused around the ZP region, which is the outer layer surrounding the embryo. For instance, in Figure \ref{fig:ann1}, despite the image depicting a real embryo, the embryologists could not clearly discern the boundary of the ZP, resulting in uncertainty about the image's authenticity. Figures \ref{fig:ann2} and \ref{fig:ann3} display blastocyst-stage embryos. Figure \ref{fig:ann2} shows a real image that was mistakenly judged as fake because the trophectoderm—the outer cell layer that forms the placenta—was poorly visible due to darkness in the image, which misled the embryologists. In contrast, Figure \ref{fig:ann3} is a synthetic image correctly identified as fake, as it lacked the end of the trophectoderm, a clear indicator for embryologists. Figures \ref{fig:ann4} and \ref{fig:ann5} show sperm sitting on ZP with abnormal morphological structure leading embryologists to identify images as synthetic. To investigate the root cause, further training of our models on embryo images with unfused sperm on the ZP is recommended followed by an evaluation of the generated synthetic data.

The participants also provided several comments, with some focusing on the presence of unusual dots within the embryos. All images containing these dots were indeed synthetic, generated by StyleGAN. One such image, Figure \ref{fig:ann6}, was specifically highlighted due to a comment on these dots. The embryologist who identified the spots mentioned having previous experience with generative models, albeit not specifically with embryos, and recognized similar artifacts in other AI-generated images. This familiarity allowed him to recognize the dots as a sign of synthetic generation. In contrast, embryologists who were less familiar with synthetic images did not observe these abnormalities. Another remark was that some cells in the StyleGAN-generated images appeared flat, as seen in Figure \ref{fig:ann6}. This contributed to the perception that these images were synthetic as real embryo cells are rounder, not flat.

\section*{Discussion}
The primary goal of this study was to assess whether the use of synthetic data can address embryo data scarcity and improve classification accuracy, with a particular emphasis on ensuring data quality. Our findings confirm that this approach is indeed effective: training classifiers solely on synthetic data resulted in good accuracy, and incorporating synthetic images alongside real ones further slightly improved performance. Notably, this performance boost persisted even when tested on an external dataset, demonstrating the potential of synthetic data for enhancing model generalization.

\begin{figure}[htb!]
  \centering
    \begin{subfigure}{0.224\columnwidth}
    \includegraphics[width=\textwidth]{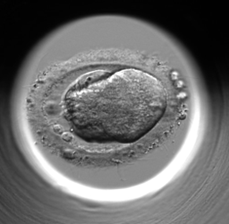} 
    \caption{2-cell \textcolor{red}{(Morula)}}
    \label{fig:fig8a}
    \end{subfigure}
    \begin{subfigure}{0.22\columnwidth}
    \includegraphics[width=\textwidth]{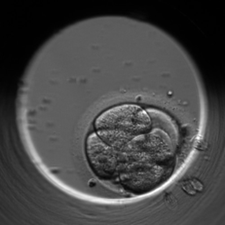} 
    \caption{4-cell \textcolor{red}{(8-cell)}}
    \label{fig:fig8b}
    \end{subfigure}
    \begin{subfigure}{0.22\columnwidth}
    \includegraphics[width=\textwidth]{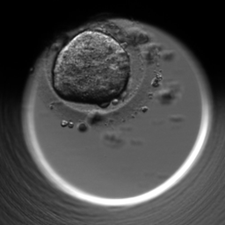} 
    \caption{Morula \textcolor{red}{(2-cell)}}
    \label{fig:fig8c}
    \end{subfigure}
    \begin{subfigure}{0.221\columnwidth}
    \includegraphics[width=\textwidth]{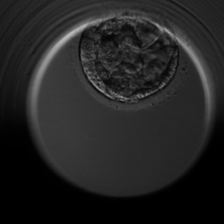} 
    \caption{Blastocyst \textcolor{red}{(Morula)}}
    \label{fig:fig8d}
    \end{subfigure}
\caption{Examples of embryos misclassified by the real-data-only model but correctly classified with real and synthetic data. True (black) and misclassified (red) classes are shown below each image.}
  \label{fig:fig8}
\end{figure}

To understand how synthetic images improved model performance, we examined samples misclassified by the real-data-only model but correctly classified when using both real and synthetic data. Figure \ref{fig:fig8} shows such embryo images, with the true class and the real-data-only misclassification noted below each image. In the first image, misclassification may stem from an indistinct membrane, resembling a morulated embryo. The second image could reflect fragmentation being mistaken for individual cells. In the third, a central structure within the morula may have been misidentified as a membrane. The fourth image shows an embryo in early blastocyst formation, indicated by a small fluid cavity, rather than a fully developed blastocyst.

We trained separate generative models for each stage, rather than using a single conditional model, to produce higher-quality synthetic images. This decision was guided by empirical results, where separate LDM models outperformed the conditional approach in both FID scores and visual quality. However, this comparison was limited to LDMs, not GANs, which we acknowledge as a limitation. To evaluate the impact of these synthetic images on classification, we tested three models—VGG, ResNet, and ViT—observing distinct performance patterns. VGG achieved the best overall results, likely due to its simple architecture and use of small 3×3 kernels, which effectively capture local features in embryo images. In contrast, ResNet and ViT performed less favorably, particularly ViT when trained from scratch, as transformers typically require larger datasets to perform well. The more complex architectures of ResNet and ViT, which excel at capturing global dependencies, may better demonstrate their strengths with larger training datasets.

Furthermore, the synthetic images were convincing enough to lead embryologists to believe they were real, highlighting the high quality of the generated data. We also found that images generated by the LDM exhibit higher fidelity compared to those produced by the GAN. This superiority was supported by our quantitative results, including lower FID scores and higher classification accuracies on the test data, as well as qualitative results from expert evaluations, where the LDM images fooled embryologists more frequently. Despite these positive outcomes, certain challenges were identified during the evaluation process.

One such challenge noted by embryologists was that some images appeared too dark, making it difficult to evaluate the embryo morphology and rendering them unreliable for accurate assessment. Interestingly, this issue was not exclusive to the synthetic images: twelve comments referred to the real images, while eight concerned the LDM-generated images. This occasional darkness in real images may stem from a bubble obstructing the camera’s direct view, floating between the camera and the EmbryoSlide well containing the embryo. This suggests that the LDM model learned and reproduced this characteristic from the real dataset. A possible solution to address this issue would be to preprocess the real images before training the generative models. There could be a quality control step, where only images meeting specific brightness and contrast thresholds are included in the training dataset. Very dark images could either be filtered out or adjusted using brightness correction techniques, ensuring that structural features essential for embryo assessment remain unaffected. This approach would ensure that the generative models are exposed only to high-quality examples, reducing the likelihood of producing too dark synthetic images. Addressing such data quality issues is crucial for further improving the effectiveness of the models.

In addition to these considerations, while our work demonstrates the potential of generative models to enhance embryo classification and overcome data scarcity, several limitations should be acknowledged. We did not assess the similarity between synthetic and real images, which could be measured using similarity metrics to rule out the risk of overfitting. Overfitting is a common problem when dealing with small datasets and powerful deep-learning models. It happens when models achieve very good results on the training data but struggle with new data (such as test data) because they have memorized specific patterns rather than learned to generalize effectively. In our case, overfitting would mean that the generative models ‘copied’ the real images rather than generated new, diverse ones. Since incorporating synthetic data led to improved classification performance, even on an external dataset, we believe it is unlikely that overfitting occurred. The improved performance suggests that the synthetic data adds value rather than merely replicating the training set. However, to fully rule out overfitting or mode collapse in the generative models, future work should quantify training-generated image similarity using appropriate metrics. Additionally, validating the classification models on embryo images from all developmental stages, sourced from different clinics and imaging systems, will be essential for assessing true generalization and ensuring the models do not overfit to the distribution of synthetic data from a single source.

Another limitation of this study is the selection of cell stages included in the dataset. We focused on the 2-, 4-, 8-cell, morula, and blastocyst stages while excluding intermediate stages, such as the 3- and 5-cell stages, which are clinically relevant despite being less frequent. These stages are often underrepresented because they typically last less than two hours, making them more challenging to capture. However, their scarcity in datasets makes them particularly valuable in the context of missing data. Additionally, our dataset primarily included embryos with minimal fragmentation. In clinical practice, embryos with higher fragmentation are common, making classification more challenging. Focusing on lower-fragmentation embryos may have simplified the task, potentially overestimating our model’s performance. Future work should address these limitations by incorporating intermediate cell stages, such as 3- and 5-cell embryos, as well as embryos with higher fragmentation rates. Furthermore, expanding the dataset to include multiple focal planes would better capture the full spectrum of embryo variations observed in clinical practice, ultimately improving model robustness and clinical applicability.

Furthermore, embryologists analyzing the synthetic images were all from the same clinic, Volvat Spiren, where they are accustomed to viewing and analyzing embryo images from their lab only. Upon analyzing the qualitative results, we found that certain image regions, such as the dark background or the embryo’s zona pellucida (ZP), sometimes misled embryologists into identifying real images as synthetic. These images were from the second dataset we used. Although images from this second dataset were captured with the same Embryoscope™, variations in lab settings may have impacted image quality, introducing potential bias into the evaluation process. To address this, future research could build on this work by involving a more diverse group of embryologists from different clinics to refine the process of embryo synthetic image evaluation. Another potential direction for future research could involve incorporating images of embryos undergoing necrosis into model training. Our real dataset contained a limited number of such images, which impacted the generative models’ ability to produce high-fidelity embryo images under necrosis. Introducing synthetic embryo images under necrosis could prove valuable for future studies by enabling models to better recognize when an embryo is no longer viable. This advancement would aid embryologists in determining when an embryo is unsuitable for use, enhancing decision-making in clinical settings.

\section*{Code Availability}
All the code used in this study is publicly available in our GitHub repository \url{https://github.com/orianapresacan/EmbryoStageGen} to ensure reproducibility. Additionally, the repository includes links to the checkpoints of the best-performing models, which were used for image generation and classification tasks. 

\section*{Data Availability}
The two real datasets used in this study are publicly available and can be accessed on Zenodo at \url{https://zenodo.org/records/6390798} and \url{https://zenodo.org/records/14253170}. The synthetic dataset we generated and used to train the classification models can be found at \url{https://huggingface.co/datasets/deepsynthbody/synembryo_latentdiffusion} and \url{https://huggingface.co/datasets/deepsynthbody/synembryo_stylegan}. The blastocyst dataset used for evaluation is available at \url{https://github.com/ih-lab/STORK}.

\bibliography{sample}

\section*{Author contributions statement}
O.P., A.S., V.T., A.D., and M.R. conceived the experiments and study design. O.P. conducted the experiments. A.S. conducted the background research on embryology and AI. M.S. and M.I. collected the data, while A.S. was responsible for the preprocessing. A.D. developed the web application. M.S. and M.I. performed qualitative evaluations of the synthetic images. O.P., V.T., M.R., A.A., and A.S. analyzed the results. All authors reviewed and contributed to the final manuscript.

\end{document}


\maketitle
\renewcommand{\thetable}{S\arabic{table}}
\begin{table}[htb!]
\centering
\small
\scalebox{0.9}{
\begin{tabular}{|cccccc|}
\hline
\multicolumn{6}{|c|}{\multirow{2}{*}{\textbf{FID - StyleGAN (transfer learning)}}}  \\
\multicolumn{6}{|c|}{}                                                               \\ \hline
\multicolumn{1}{|c|}{kimg/dataset}  & \multicolumn{1}{c|}{\textbf{2-cell}} & \multicolumn{1}{c|}{\textbf{4-cell}} & \multicolumn{1}{c|}{\textbf{8-cell}} & \multicolumn{1}{c|}{\textbf{blastocyst}} & \textbf{morula} \\ \hline
\multicolumn{1}{|c|}{\textbf{50}}   & \multicolumn{1}{c|}{90}             & \multicolumn{1}{c|}{67}             & \multicolumn{1}{c|}{95}             & \multicolumn{1}{c|}{97}                  & 91              \\ \hline
\multicolumn{1}{|c|}{\textbf{100}}  & \multicolumn{1}{c|}{52}             & \multicolumn{1}{c|}{40}             & \multicolumn{1}{c|}{64}             & \multicolumn{1}{c|}{68}                  & 65              \\ \hline
\multicolumn{1}{|c|}{\textbf{150}}  & \multicolumn{1}{c|}{38}             & \multicolumn{1}{c|}{35}             & \multicolumn{1}{c|}{54}             & \multicolumn{1}{c|}{58}                  & 60              \\ \hline
\multicolumn{1}{|c|}{\textbf{200}}  & \multicolumn{1}{c|}{37}             & \multicolumn{1}{c|}{31}             & \multicolumn{1}{c|}{48}             & \multicolumn{1}{c|}{54}                  & 52              \\ \hline
\multicolumn{1}{|c|}{\textbf{250}}  & \multicolumn{1}{c|}{34}             & \multicolumn{1}{c|}{29}             & \multicolumn{1}{c|}{45}             & \multicolumn{1}{c|}{53}                  & 46              \\ \hline
\multicolumn{1}{|c|}{\textbf{300}}  & \multicolumn{1}{c|}{35}             & \multicolumn{1}{c|}{26}             & \multicolumn{1}{c|}{45}             & \multicolumn{1}{c|}{52}                  & 45              \\ \hline
\multicolumn{1}{|c|}{\textbf{350}}  & \multicolumn{1}{c|}{33}             & \multicolumn{1}{c|}{25}             & \multicolumn{1}{c|}{44}             & \multicolumn{1}{c|}{50}                  & 43              \\ \hline
\multicolumn{1}{|c|}{\textbf{400}}  & \multicolumn{1}{c|}{30}             & \multicolumn{1}{c|}{28}             & \multicolumn{1}{c|}{42}             & \multicolumn{1}{c|}{49}                  & 47              \\ \hline
\multicolumn{1}{|c|}{\textbf{450}}  & \multicolumn{1}{c|}{28}             & \multicolumn{1}{c|}{29}             & \multicolumn{1}{c|}{41}             & \multicolumn{1}{c|}{47}                  & 41              \\ \hline
\multicolumn{1}{|c|}{\textbf{500}}  & \multicolumn{1}{c|}{30}             & \multicolumn{1}{c|}{26}             & \multicolumn{1}{c|}{40}             & \multicolumn{1}{c|}{44}                  & 41              \\ \hline
\multicolumn{1}{|c|}{\textbf{550}}  & \multicolumn{1}{c|}{28}             & \multicolumn{1}{c|}{27}             & \multicolumn{1}{c|}{38}             & \multicolumn{1}{c|}{43}                  & 41              \\ \hline
\multicolumn{1}{|c|}{\textbf{600}}  & \multicolumn{1}{c|}{28}             & \multicolumn{1}{c|}{26}             & \multicolumn{1}{c|}{38}             & \multicolumn{1}{c|}{49}                  & 37              \\ \hline
\multicolumn{1}{|c|}{\textbf{650}}  & \multicolumn{1}{c|}{26}             & \multicolumn{1}{c|}{25}             & \multicolumn{1}{c|}{38}             & \multicolumn{1}{c|}{48}                  & 41              \\ \hline
\multicolumn{1}{|c|}{\textbf{700}}  & \multicolumn{1}{c|}{25}             & \multicolumn{1}{c|}{25}             & \multicolumn{1}{c|}{38}             & \multicolumn{1}{c|}{46}                  & 38              \\ \hline
\multicolumn{1}{|c|}{\textbf{750}}  & \multicolumn{1}{c|}{28}             & \multicolumn{1}{c|}{25}             & \multicolumn{1}{c|}{37}             & \multicolumn{1}{c|}{45}                  & \textbf{36}     \\ \hline
\multicolumn{1}{|c|}{\textbf{800}}  & \multicolumn{1}{c|}{28}             & \multicolumn{1}{c|}{28}             & \multicolumn{1}{c|}{37}             & \multicolumn{1}{c|}{43}                  & 39              \\ \hline
\multicolumn{1}{|c|}{\textbf{850}}  & \multicolumn{1}{c|}{28}             & \multicolumn{1}{c|}{26}             & \multicolumn{1}{c|}{35}             & \multicolumn{1}{c|}{50}                  & 38              \\ \hline
\multicolumn{1}{|c|}{\textbf{900}}  & \multicolumn{1}{c|}{29}             & \multicolumn{1}{c|}{\textbf{24}}    & \multicolumn{1}{c|}{38}             & \multicolumn{1}{c|}{47}                  & 37              \\ \hline
\multicolumn{1}{|c|}{\textbf{950}}  & \multicolumn{1}{c|}{27}             & \multicolumn{1}{c|}{25}             & \multicolumn{1}{c|}{38}             & \multicolumn{1}{c|}{44}                  & 41              \\ \hline
\multicolumn{1}{|c|}{\textbf{1000}} & \multicolumn{1}{c|}{29}             & \multicolumn{1}{c|}{26}             & \multicolumn{1}{c|}{35}             & \multicolumn{1}{c|}{49}                  & 41              \\ \hline
\multicolumn{1}{|c|}{\textbf{1050}} & \multicolumn{1}{c|}{\textbf{24}}    & \multicolumn{1}{c|}{26}             & \multicolumn{1}{c|}{37}             & \multicolumn{1}{c|}{46}                  & 39              \\ \hline
\multicolumn{1}{|c|}{\textbf{1100}} & \multicolumn{1}{c|}{27}             & \multicolumn{1}{c|}{25}             & \multicolumn{1}{c|}{36}             & \multicolumn{1}{c|}{47}                  & 38              \\ \hline
\multicolumn{1}{|c|}{\textbf{1150}} & \multicolumn{1}{c|}{27}             & \multicolumn{1}{c|}{\textbf{24}}    & \multicolumn{1}{c|}{36}             & \multicolumn{1}{c|}{45}                  & 37              \\ \hline
\multicolumn{1}{|c|}{\textbf{1200}} & \multicolumn{1}{c|}{\textbf{24}}    & \multicolumn{1}{c|}{28}             & \multicolumn{1}{c|}{40}             & \multicolumn{1}{c|}{45}                  & 37              \\ \hline
\multicolumn{1}{|c|}{\textbf{1250}} & \multicolumn{1}{c|}{27}             & \multicolumn{1}{c|}{28}             & \multicolumn{1}{c|}{35}             & \multicolumn{1}{c|}{45}                  & 37              \\ \hline
\multicolumn{1}{|c|}{\textbf{1300}} & \multicolumn{1}{c|}{29}             & \multicolumn{1}{c|}{27}             & \multicolumn{1}{c|}{38}             & \multicolumn{1}{c|}{48}                  & 41              \\ \hline
\multicolumn{1}{|c|}{\textbf{1350}} & \multicolumn{1}{c|}{\textbf{24}}    & \multicolumn{1}{c|}{28}             & \multicolumn{1}{c|}{39}             & \multicolumn{1}{c|}{44}                  & 37              \\ \hline
\multicolumn{1}{|c|}{\textbf{1400}} & \multicolumn{1}{c|}{26}             & \multicolumn{1}{c|}{29}             & \multicolumn{1}{c|}{39}             & \multicolumn{1}{c|}{\textbf{41}}         & 39              \\ \hline
\multicolumn{1}{|c|}{\textbf{1450}} & \multicolumn{1}{c|}{25}             & \multicolumn{1}{c|}{26}             & \multicolumn{1}{c|}{36}             & \multicolumn{1}{c|}{44}                  & 39              \\ \hline
\multicolumn{1}{|c|}{\textbf{1500}} & \multicolumn{1}{c|}{\textbf{24}}    & \multicolumn{1}{c|}{25}             & \multicolumn{1}{c|}{\textbf{34}}    & \multicolumn{1}{c|}{44}                  & 37              \\ \hline
\end{tabular}}
\caption{FID scores for the StyleGAN model across different training epochs for the five class categories: 2-cell, 4-cell, 8-cell, morula, and blastocyst.}
\label{tab:GAN}
\end{table}

\begin{table}[htb!]
\centering
\small
\scalebox{0.9}{
\begin{tabular}{|cccccc|}
\hline
\multicolumn{6}{|c|}{\multirow{2}{*}{\textbf{FID - LDM}}} \\
\multicolumn{6}{|c|}{}                                                               \\ \hline
\multicolumn{1}{|c|}{epoch/dataset} & \multicolumn{1}{c|}{\textbf{2-cell}} & \multicolumn{1}{c|}{\textbf{4-cell}} & \multicolumn{1}{c|}{\textbf{8-cell}} & \multicolumn{1}{c|}{\textbf{blastocyst}} & \textbf{morula} \\ \hline
\multicolumn{1}{|c|}{\textbf{10}}   & \multicolumn{1}{c|}{87}             & \multicolumn{1}{c|}{98}             & \multicolumn{1}{c|}{93}             & \multicolumn{1}{c|}{76}                  & 69              \\ \hline
\multicolumn{1}{|c|}{\textbf{50}}   & \multicolumn{1}{c|}{34}             & \multicolumn{1}{c|}{37}             & \multicolumn{1}{c|}{38}             & \multicolumn{1}{c|}{31}                  & 37              \\ \hline
\multicolumn{1}{|c|}{\textbf{100}}  & \multicolumn{1}{c|}{26}             & \multicolumn{1}{c|}{\textbf{28}}    & \multicolumn{1}{c|}{33}             & \multicolumn{1}{c|}{34}                  & 35              \\ \hline
\multicolumn{1}{|c|}{\textbf{150}}  & \multicolumn{1}{c|}{\textbf{24}}    & \multicolumn{1}{c|}{29}             & \multicolumn{1}{c|}{\textbf{31}}    & \multicolumn{1}{c|}{54}                  & 47              \\ \hline
\multicolumn{1}{|c|}{\textbf{200}}  & \multicolumn{1}{c|}{35}             & \multicolumn{1}{c|}{36}             & \multicolumn{1}{c|}{33}             & \multicolumn{1}{c|}{69}                  & 65              \\ \hline
\multicolumn{1}{|c|}{\textbf{250}}  & \multicolumn{1}{c|}{55}             & \multicolumn{1}{c|}{54}             & \multicolumn{1}{c|}{38}             & \multicolumn{1}{c|}{61}                  & 73              \\ \hline
\multicolumn{1}{|c|}{\textbf{300}}  & \multicolumn{1}{c|}{84}             & \multicolumn{1}{c|}{83}             & \multicolumn{1}{c|}{51}             & \multicolumn{1}{c|}{47}                  & 73              \\ \hline
\multicolumn{1}{|c|}{\textbf{350}}  & \multicolumn{1}{c|}{104}            & \multicolumn{1}{c|}{105}            & \multicolumn{1}{c|}{69}             & \multicolumn{1}{c|}{35}                  & 65              \\ \hline
\multicolumn{1}{|c|}{\textbf{400}}  & \multicolumn{1}{c|}{114}            & \multicolumn{1}{c|}{118}            & \multicolumn{1}{c|}{81}             & \multicolumn{1}{c|}{27}                  & 59              \\ \hline
\multicolumn{1}{|c|}{\textbf{450}}  & \multicolumn{1}{c|}{119}            & \multicolumn{1}{c|}{124}            & \multicolumn{1}{c|}{85}             & \multicolumn{1}{c|}{23}                  & 49              \\ \hline
\multicolumn{1}{|c|}{\textbf{500}}  & \multicolumn{1}{c|}{118}            & \multicolumn{1}{c|}{126}            & \multicolumn{1}{c|}{89}             & \multicolumn{1}{c|}{19}                  & 45              \\ \hline
\multicolumn{1}{|c|}{\textbf{550}}  & \multicolumn{1}{c|}{119}            & \multicolumn{1}{c|}{123}            & \multicolumn{1}{c|}{89}             & \multicolumn{1}{c|}{16}                  & 37              \\ \hline
\multicolumn{1}{|c|}{\textbf{600}}  & \multicolumn{1}{c|}{117}            & \multicolumn{1}{c|}{117}            & \multicolumn{1}{c|}{85}             & \multicolumn{1}{c|}{15}                  & 33              \\ \hline
\multicolumn{1}{|c|}{\textbf{650}}  & \multicolumn{1}{c|}{116}            & \multicolumn{1}{c|}{114}            & \multicolumn{1}{c|}{85}             & \multicolumn{1}{c|}{13}                  & 27              \\ \hline
\multicolumn{1}{|c|}{\textbf{700}}  & \multicolumn{1}{c|}{110}            & \multicolumn{1}{c|}{103}            & \multicolumn{1}{c|}{80}             & \multicolumn{1}{c|}{12}                  & 24              \\ \hline
\multicolumn{1}{|c|}{\textbf{750}}  & \multicolumn{1}{c|}{110}            & \multicolumn{1}{c|}{96}             & \multicolumn{1}{c|}{74}             & \multicolumn{1}{c|}{11}                  & 21              \\ \hline
\multicolumn{1}{|c|}{\textbf{800}}  & \multicolumn{1}{c|}{106}            & \multicolumn{1}{c|}{92}             & \multicolumn{1}{c|}{72}             & \multicolumn{1}{c|}{11}                  & 19              \\ \hline
\multicolumn{1}{|c|}{\textbf{850}}  & \multicolumn{1}{c|}{99}             & \multicolumn{1}{c|}{82}             & \multicolumn{1}{c|}{60}             & \multicolumn{1}{c|}{11}                  & 18              \\ \hline
\multicolumn{1}{|c|}{\textbf{900}}  & \multicolumn{1}{c|}{94}             & \multicolumn{1}{c|}{74}             & \multicolumn{1}{c|}{60}             & \multicolumn{1}{c|}{11}                  & 17              \\ \hline
\multicolumn{1}{|c|}{\textbf{950}}  & \multicolumn{1}{c|}{91}             & \multicolumn{1}{c|}{68}             & \multicolumn{1}{c|}{53}             & \multicolumn{1}{c|}{11}                  & 14              \\ \hline
\multicolumn{1}{|c|}{\textbf{1000}} & \multicolumn{1}{c|}{86}             & \multicolumn{1}{c|}{61}             & \multicolumn{1}{c|}{50}             & \multicolumn{1}{c|}{\textbf{10}}         & \textbf{14}     \\ \hline
\end{tabular}}
\caption{FID scores for the LDM across different training epochs for the five class categories: 2-cell, 4-cell, 8-cell, morula, and blastocyst.}
\label{tab:LDM}
\end{table}
\begin{table}[ht]
\centering
\resizebox{\textwidth}{!}{%
\begin{tabular}{|c|c|c|c|c|c|c|c|c|}
\hline
\textbf{Real} & \textbf{StyleGAN} & \textbf{LDM} & \textbf{Accuracy} & \textbf{F1 Score} & \textbf{Precision} & \textbf{Recall} & \textbf{MCC} \\
\hline
0 & 0    & 500  & 70.68 ± 5.32 & 0.716 ± 0.021 & 0.717 ± 0.053 & 0.713 ± 0.055 & 0.664 ± 0.052 \\ \hline
0 & 500  & 0    & 47.36 ± 1.70 & 0.351 ± 0.098 & 0.474 ± 0.017 & 0.358 ± 0.017 & 0.380 ± 0.024 \\\hline
0 & 250  & 250  & 74.24 ± 2.78 & 0.789 ± 0.026 & 0.742 ± 0.028 & 0.735 ± 0.031 & 0.689 ± 0.035 \\\hline
0 & 500  & 500  & 84.20 ± 1.62 & 0.866 ± 0.011 & 0.842 ± 0.016 & 0.839 ± 0.017 & 0.809 ± 0.019 \\\hline
0 & 1000 & 1000 & 88.44 ± 1.10 & 0.892 ± 0.007 & 0.884 ± 0.011 & 0.883 ± 0.012 & 0.859 ± 0.014 \\\hline
0 & 2000 & 2000 & 91.48 ± 1.01 & 0.920 ± 0.008 & 0.915 ± 0.010 & 0.915 ± 0.010 & 0.897 ± 0.014 \\\hline
0 & 3000 & 3000 & 92.16 ± 1.43 & 0.926 ± 0.011 & 0.922 ± 0.014 & 0.920 ± 0.016 & 0.903 ± 0.016 \\\hline
0 & 4000 & 4000 & 93.60 ± 1.45 & 0.939 ± 0.012 & 0.936 ± 0.014 & 0.936 ± 0.014 & 0.920 ± 0.017 \\\hline
0 & 5000 & 5000 & 93.28 ± 0.72 & 0.936 ± 0.005 & 0.933 ± 0.007 & 0.932 ± 0.007 & 0.916 ± 0.007 \\\hline
1000 & 0    & 0    & 92.76 ± 0.89 & 0.930 ± 0.007 & 0.928 ± 0.009 & 0.927 ± 0.009 & 0.909 ± 0.009 \\\hline
1000 & 500  & 500  & 93.84 ± 0.52 & 0.938 ± 0.003 & 0.937 ± 0.003 & 0.937 ± 0.003 & 0.923 ± 0.005 \\\hline
1000 & 1000 & 1000 & 95.08 ± 0.52 & 0.951 ± 0.005 & 0.950 ± 0.005 & 0.950 ± 0.005 & 0.938 ± 0.006 \\\hline
1000 & 2000 & 2000 & 95.08 ± 0.70 & 0.951 ± 0.008 & 0.950 ± 0.008 & 0.950 ± 0.008 & 0.937 ± 0.011 \\\hline
1000 & 3000 & 3000 & 95.24 ± 0.65 & 0.952 ± 0.007 & 0.952 ± 0.007 & 0.952 ± 0.007 & 0.939 ± 0.011 \\\hline
1000 & 4000 & 4000 & 95.56 ± 0.46 & 0.955 ± 0.008 & 0.954 ± 0.008 & 0.954 ± 0.008 & 0.944 ± 0.008 \\\hline
1000 & 5000 & 5000 & 95.20 ± 0.53 & 0.951 ± 0.004 & 0.950 ± 0.004 & 0.950 ± 0.004 & 0.939 ± 0.005 \\\hline

\end{tabular}%
}
\caption{Mean performance metrics (± standard deviation) of VGG16 models on the test dataset (consisting of real images), when trained from scratch using various combinations of real and synthetic images.}
\end{table}

\begin{table}[ht]
\centering
\resizebox{\textwidth}{!}{%
\begin{tabular}{|c|c|c|c|c|c|c|c|c|}
\hline
\textbf{Real} & \textbf{StyleGAN} & \textbf{LDM} & \textbf{Accuracy} & \textbf{F1 Score} & \textbf{Precision} & \textbf{Recall} & \textbf{MCC} \\
\hline
0 & 0    & 500  & 84.8 ± 5.04   & 0.831 ± 0.070 & 0.872 ± 0.030 & 0.848 ± 0.050 & 0.821 ± 0.057 \\\hline
0 & 500  & 0    & 44.26 ± 7.94  & 0.327 ± 0.123 & 0.395 ± 0.215 & 0.442 ± 0.079 & 0.349 ± 0.087 \\\hline
0 & 250  & 250  & 88.13 ± 2.20  & 0.881 ± 0.021 & 0.886 ± 0.017 & 0.881 ± 0.022 & 0.852 ± 0.026 \\\hline
0 & 500  & 500  & 86.2 ± 3.41   & 0.862 ± 0.042 & 0.874 ± 0.033 & 0.865 ± 0.039 & 0.832 ± 0.042 \\\hline
0 & 1000 & 1000 & 91.13 ± 2.31  & 0.911 ± 0.022 & 0.916 ± 0.019 & 0.911 ± 0.023 & 0.892 ± 0.026 \\\hline
0 & 2000 & 2000 & 92.16 ± 1.15  & 0.928 ± 0.012 & 0.931 ± 0.011 & 0.928 ± 0.012 & 0.912 ± 0.013 \\\hline
0 & 3000 & 3000 & 90 ± 2.20     & 0.899 ± 0.021 & 0.902 ± 0.017 & 0.899 ± 0.021 & 0.876 ± 0.027 \\\hline
0 & 4000 & 4000 & 92 ± 1.31     & 0.919 ± 0.013 & 0.922 ± 0.014 & 0.919 ± 0.013 & 0.901 ± 0.015 \\\hline
0 & 5000 & 5000 & 92.8 ± 1.5    & 0.932 ± 0.017 & 0.935 ± 0.008 & 0.935 ± 0.020 & 0.916 ± 0.020 \\\hline
1000 & 0    & 0    & 94.46 ± 0.64  & 0.944 ± 0.006 & 0.944 ± 0.013 & 0.944 ± 0.006 & 0.934 ± 0.007 \\\hline
1000 & 500  & 500  & 96.46 ± 0.41  & 0.961 ± 0.003 & 0.964 ± 0.006 & 0.962 ± 0.003 & 0.956 ± 0.005 \\\hline
1000 & 1000 & 1000 & 95.66 ± 0.23  & 0.956 ± 0.002 & 0.954 ± 0.003 & 0.954 ± 0.002 & 0.943 ± 0.003 \\\hline
1000 & 2000 & 2000 & 96.33 ± 0.75  & 0.962 ± 0.006 & 0.962 ± 0.004 & 0.962 ± 0.006 & 0.952 ± 0.006 \\\hline
1000 & 3000 & 3000 & 96 ± 1.5      & 0.957 ± 0.016 & 0.957 ± 0.016 & 0.957 ± 0.016 & 0.947 ± 0.019 \\\hline
1000 & 4000 & 4000 & 96.26 ± 1.1   & 0.961 ± 0.009 & 0.962 ± 0.008 & 0.961 ± 0.009 & 0.951 ± 0.009 \\\hline
1000 & 5000 & 5000 & 96.96 ± 0.3   & 0.964 ± 0.004 & 0.964 ± 0.005 & 0.964 ± 0.005 & 0.955 ± 0.006 \\
\hline
\end{tabular}%
}
\caption{Mean performance metrics (± standard deviation) of VGG16 models on the test dataset (consisting of real images), when fine-tuned using various combinations of real and synthetic images.}
\end{table}

\begin{table}[htb!]
\centering

\begin{adjustbox}{width=\textwidth}
\begin{tabular}{ccccccccccccc}
\hline
\multicolumn{3}{|c|}{\textbf{Training Data}} & \multicolumn{10}{c|}{\textbf{ResNet50}} \\ \hline
\multicolumn{1}{|c|}{\multirow{2}{*}{\textbf{Real}}} & \multicolumn{1}{c|}{\multirow{2}{*}{\textbf{StyleGAN}}} & \multicolumn{1}{c|}{\multirow{2}{*}{\textbf{LDM}}} & \multicolumn{5}{c|}{\textbf{from scratch}}                          & \multicolumn{5}{c|}{\textbf{pretrained}}                                          \\ \cline{4-13} 
\multicolumn{1}{|c|}{}                               & \multicolumn{1}{c|}{}                                   & \multicolumn{1}{c|}{}                              & \multicolumn{1}{c|}{\textbf{Accuracy}} & \multicolumn{1}{c|}{\textbf{F1 Score}} & \multicolumn{1}{c|}{\textbf{Precision}} & \multicolumn{1}{c|}{\textbf{Recall}} & \multicolumn{1}{c|}{\textbf{MCC}}  & \multicolumn{1}{c|}{\textbf{Accuracy}} & \multicolumn{1}{c|}{\textbf{F1 Score}} & \multicolumn{1}{c|}{\textbf{Precision}} & \multicolumn{1}{c|}{\textbf{Recall}} & \multicolumn{1}{c|}{\textbf{MCC}}  \\ \hline
\multicolumn{1}{|c|}{0}                              & \multicolumn{1}{c|}{0}                                  & \multicolumn{1}{c|}{500}                           & \multicolumn{1}{c|}{44\%}              & \multicolumn{1}{c|}{0.42}              & \multicolumn{1}{c|}{0.54}               & \multicolumn{1}{c|}{0.44}            & \multicolumn{1}{c|}{0.33}          & \multicolumn{1}{c|}{75\%}              & \multicolumn{1}{c|}{0.71}              & \multicolumn{1}{c|}{0.81}               & \multicolumn{1}{c|}{0.75}            & \multicolumn{1}{c|}{0.71}          \\ \hline
\multicolumn{1}{|c|}{0}                              & \multicolumn{1}{c|}{500}                                & \multicolumn{1}{c|}{0}                             & \multicolumn{1}{c|}{38\%}              & \multicolumn{1}{c|}{0.28}              & \multicolumn{1}{c|}{0.42}               & \multicolumn{1}{c|}{0.38}            & \multicolumn{1}{c|}{0.25}          & \multicolumn{1}{c|}{44\%}              & \multicolumn{1}{c|}{0.32}              & \multicolumn{1}{c|}{0.55}               & \multicolumn{1}{c|}{0.44}            & \multicolumn{1}{c|}{0.37}          \\ \hline
\multicolumn{1}{|c|}{0}                              & \multicolumn{1}{c|}{250}                                & \multicolumn{1}{c|}{250}                           & \multicolumn{1}{c|}{45\%}              & \multicolumn{1}{c|}{0.44}              & \multicolumn{1}{c|}{0.55}               & \multicolumn{1}{c|}{0.45}            & \multicolumn{1}{c|}{0.32}          & \multicolumn{1}{c|}{86\%}              & \multicolumn{1}{c|}{0.86}              & \multicolumn{1}{c|}{0.88}               & \multicolumn{1}{c|}{0.86}            & \multicolumn{1}{c|}{0.83}          \\ \hline
\multicolumn{1}{|c|}{0}                              & \multicolumn{1}{c|}{500}                                & \multicolumn{1}{c|}{500}                           & \multicolumn{1}{c|}{55\%}              & \multicolumn{1}{c|}{0.53}              & \multicolumn{1}{c|}{0.64}               & \multicolumn{1}{c|}{0.55}            & \multicolumn{1}{c|}{0.46}          & \multicolumn{1}{c|}{83\%}              & \multicolumn{1}{c|}{0.83}              & \multicolumn{1}{c|}{0.85}               & \multicolumn{1}{c|}{0.83}            & \multicolumn{1}{c|}{0.79}          \\ \hline
\multicolumn{1}{|c|}{0}                              & \multicolumn{1}{c|}{1000}                               & \multicolumn{1}{c|}{1000}                          & \multicolumn{1}{c|}{74\%}              & \multicolumn{1}{c|}{0.72}              & \multicolumn{1}{c|}{0.78}               & \multicolumn{1}{c|}{0.74}            & \multicolumn{1}{c|}{0.68}          & \multicolumn{1}{c|}{\textbf{89\%}}     & \multicolumn{1}{c|}{\textbf{0.89}}     & \multicolumn{1}{c|}{\textbf{0.9}}       & \multicolumn{1}{c|}{\textbf{0.89}}   & \multicolumn{1}{c|}{\textbf{0.87}} \\ \hline
\multicolumn{1}{|c|}{0}                              & \multicolumn{1}{c|}{2000}                               & \multicolumn{1}{c|}{2000}                          & \multicolumn{1}{c|}{80\%}              & \multicolumn{1}{c|}{0.79}              & \multicolumn{1}{c|}{0.83}               & \multicolumn{1}{c|}{0.8}             & \multicolumn{1}{c|}{0.75}          & \multicolumn{1}{c|}{88\%}              & \multicolumn{1}{c|}{0.88}              & \multicolumn{1}{c|}{0.89}               & \multicolumn{1}{c|}{0.88}            & \multicolumn{1}{c|}{0.85}          \\ \hline
\multicolumn{1}{|c|}{0}                              & \multicolumn{1}{c|}{3000}                               & \multicolumn{1}{c|}{3000}                          & \multicolumn{1}{c|}{81\%}              & \multicolumn{1}{c|}{0.81}              & \multicolumn{1}{c|}{0.84}               & \multicolumn{1}{c|}{0.81}            & \multicolumn{1}{c|}{0.77}          & \multicolumn{1}{c|}{87\%}              & \multicolumn{1}{c|}{0.87}              & \multicolumn{1}{c|}{0.9}                & \multicolumn{1}{c|}{0.87}            & \multicolumn{1}{c|}{0.85}          \\ \hline
\multicolumn{1}{|c|}{0}                              & \multicolumn{1}{c|}{4000}                               & \multicolumn{1}{c|}{4000}                          & \multicolumn{1}{c|}{85\%}              & \multicolumn{1}{c|}{0.84}              & \multicolumn{1}{c|}{0.86}               & \multicolumn{1}{c|}{0.85}            & \multicolumn{1}{c|}{0.81}          & \multicolumn{1}{c|}{88\%}              & \multicolumn{1}{c|}{0.88}              & \multicolumn{1}{c|}{0.9}                & \multicolumn{1}{c|}{0.88}            & \multicolumn{1}{c|}{0.86}          \\ \hline
\multicolumn{1}{|c|}{0}                              & \multicolumn{1}{c|}{5000}                               & \multicolumn{1}{c|}{5000}                          & \multicolumn{1}{c|}{\textbf{87\%}}     & \multicolumn{1}{c|}{\textbf{0.87}}     & \multicolumn{1}{c|}{\textbf{0.88}}      & \multicolumn{1}{c|}{\textbf{0.87}}   & \multicolumn{1}{c|}{\textbf{0.84}} & \multicolumn{1}{c|}{87\%}              & \multicolumn{1}{c|}{0.87}              & \multicolumn{1}{c|}{0.89}               & \multicolumn{1}{c|}{0.87}            & \multicolumn{1}{c|}{0.84}          \\ \hline
\multicolumn{13}{l}{}    \\ \hline
\multicolumn{1}{|c|}{1000}                           & \multicolumn{1}{c|}{0}                                  & \multicolumn{1}{c|}{0}                             & \multicolumn{1}{c|}{70\%}              & \multicolumn{1}{c|}{0.7}               & \multicolumn{1}{c|}{0.72}               & \multicolumn{1}{c|}{0.7}             & \multicolumn{1}{c|}{0.63}          & \multicolumn{1}{c|}{96\%}              & \multicolumn{1}{c|}{0.95}              & \multicolumn{1}{c|}{0.96}               & \multicolumn{1}{c|}{0.96}            & \multicolumn{1}{c|}{0.95}          \\ \hline
\multicolumn{1}{|c|}{1000}                           & \multicolumn{1}{c|}{500}                                & \multicolumn{1}{c|}{500}                           & \multicolumn{1}{c|}{76\%}              & \multicolumn{1}{c|}{0.76}              & \multicolumn{1}{c|}{0.77}               & \multicolumn{1}{c|}{0.76}            & \multicolumn{1}{c|}{0.7}           & \multicolumn{1}{c|}{96\%}              & \multicolumn{1}{c|}{0.96}              & \multicolumn{1}{c|}{0.96}               & \multicolumn{1}{c|}{0.96}            & \multicolumn{1}{c|}{0.95}          \\ \hline
\multicolumn{1}{|c|}{1000}                           & \multicolumn{1}{c|}{1000}                               & \multicolumn{1}{c|}{1000}                          & \multicolumn{1}{c|}{85\%}              & \multicolumn{1}{c|}{0.85}              & \multicolumn{1}{c|}{0.85}               & \multicolumn{1}{c|}{0.85}            & \multicolumn{1}{c|}{0.81}          & \multicolumn{1}{c|}{96\%}              & \multicolumn{1}{c|}{0.95}              & \multicolumn{1}{c|}{0.96}               & \multicolumn{1}{c|}{0.96}            & \multicolumn{1}{c|}{0.95}          \\ \hline
\multicolumn{1}{|c|}{1000}                           & \multicolumn{1}{c|}{2000}                               & \multicolumn{1}{c|}{2000}                          & \multicolumn{1}{c|}{88\%}              & \multicolumn{1}{c|}{0.88}              & \multicolumn{1}{c|}{0.88}               & \multicolumn{1}{c|}{0.88}            & \multicolumn{1}{c|}{0.85}          & \multicolumn{1}{c|}{96\%}              & \multicolumn{1}{c|}{0.96}              & \multicolumn{1}{c|}{0.96}               & \multicolumn{1}{c|}{0.96}            & \multicolumn{1}{c|}{0.95}          \\ \hline
\multicolumn{1}{|c|}{1000}                           & \multicolumn{1}{c|}{3000}                               & \multicolumn{1}{c|}{3000}                          & \multicolumn{1}{c|}{87\%}              & \multicolumn{1}{c|}{0.87}              & \multicolumn{1}{c|}{0.88}               & \multicolumn{1}{c|}{0.87}            & \multicolumn{1}{c|}{0.84}          & \multicolumn{1}{c|}{96\%}              & \multicolumn{1}{c|}{0.96}              & \multicolumn{1}{c|}{0.96}               & \multicolumn{1}{c|}{0.96}            & \multicolumn{1}{c|}{0.95}          \\ \hline
\multicolumn{1}{|c|}{1000}                           & \multicolumn{1}{c|}{4000}                               & \multicolumn{1}{c|}{4000}                          & \multicolumn{1}{c|}{90\%}              & \multicolumn{1}{c|}{0.9}               & \multicolumn{1}{c|}{0.9}                & \multicolumn{1}{c|}{0.9}             & \multicolumn{1}{c|}{0.87}          & \multicolumn{1}{c|}{\textbf{97\%}}     & \multicolumn{1}{c|}{\textbf{0.97}}     & \multicolumn{1}{c|}{\textbf{0.97}}      & \multicolumn{1}{c|}{\textbf{0.97}}   & \multicolumn{1}{c|}{\textbf{0.96}} \\ \hline
\multicolumn{1}{|c|}{1000}                           & \multicolumn{1}{c|}{5000}                               & \multicolumn{1}{c|}{5000}                          & \multicolumn{1}{c|}{\textbf{90\%}}     & \multicolumn{1}{c|}{\textbf{0.9}}      & \multicolumn{1}{c|}{\textbf{0.91}}      & \multicolumn{1}{c|}{\textbf{0.9}}    & \multicolumn{1}{c|}{\textbf{0.88}} & \multicolumn{1}{c|}{94\%}              & \multicolumn{1}{c|}{0.94}              & \multicolumn{1}{c|}{0.94}               & \multicolumn{1}{c|}{0.94}            & \multicolumn{1}{c|}{0.93}          \\ \hline
\end{tabular}
\end{adjustbox}
\caption{Classification results of the ResNet50 model on the test dataset (comprised of real images), trained using various combinations of real and synthetic images.}
\label{tab:resnet}
\end{table}
\begin{table}[htb!]
\centering
\small
\begin{adjustbox}{width=\textwidth}
\begin{tabular}{ccccccccccccc}
\hline
\multicolumn{3}{|c|}{\textbf{Training Data}}                                                                                                                        & \multicolumn{10}{c|}{\textbf{Transformer (ViT\_b\_16)}}                                                                                                                          \\ \hline
\multicolumn{1}{|c|}{\multirow{2}{*}{\textbf{Real}}} & \multicolumn{1}{c|}{\multirow{2}{*}{\textbf{StyleGAN}}} & \multicolumn{1}{c|}{\multirow{2}{*}{\textbf{LDM}}} & \multicolumn{5}{c|}{\textbf{from scratch}}                                                                                                                                                            & \multicolumn{5}{c|}{\textbf{pretrained}}                                                                                                                                                              \\ \cline{4-13} 
\multicolumn{1}{|c|}{}                               & \multicolumn{1}{c|}{}                                   & \multicolumn{1}{c|}{}                              & \multicolumn{1}{c|}{\textbf{Accuracy}} & \multicolumn{1}{c|}{\textbf{F1 Score}} & \multicolumn{1}{c|}{\textbf{Precision}} & \multicolumn{1}{c|}{\textbf{Recall}} & \multicolumn{1}{c|}{\textbf{MCC}}  & \multicolumn{1}{c|}{\textbf{Accuracy}} & \multicolumn{1}{c|}{\textbf{F1 Score}} & \multicolumn{1}{c|}{\textbf{Precision}} & \multicolumn{1}{c|}{\textbf{Recall}} & \multicolumn{1}{c|}{\textbf{MCC}}  \\ \hline
\multicolumn{1}{|c|}{0}                              & \multicolumn{1}{c|}{0}                                  & \multicolumn{1}{c|}{500}                           & \multicolumn{1}{c|}{40\%}              & \multicolumn{1}{c|}{0.38}              & \multicolumn{1}{c|}{0.42}               & \multicolumn{1}{c|}{0.4}             & \multicolumn{1}{c|}{0.26}          & \multicolumn{1}{c|}{83\%}              & \multicolumn{1}{c|}{0.83}              & \multicolumn{1}{c|}{0.84}               & \multicolumn{1}{c|}{0.83}            & \multicolumn{1}{c|}{0.8}           \\ \hline
\multicolumn{1}{|c|}{0}                              & \multicolumn{1}{c|}{500}                                & \multicolumn{1}{c|}{0}                             & \multicolumn{1}{c|}{25\%}              & \multicolumn{1}{c|}{0.13}              & \multicolumn{1}{c|}{0.11}               & \multicolumn{1}{c|}{0.25}            & \multicolumn{1}{c|}{0.11}          & \multicolumn{1}{c|}{83\%}              & \multicolumn{1}{c|}{0.82}              & \multicolumn{1}{c|}{0.83}               & \multicolumn{1}{c|}{0.83}            & \multicolumn{1}{c|}{0.78}          \\ \hline
\multicolumn{1}{|c|}{0}                              & \multicolumn{1}{c|}{250}                                & \multicolumn{1}{c|}{250}                           & \multicolumn{1}{c|}{36\%}              & \multicolumn{1}{c|}{0.34}              & \multicolumn{1}{c|}{0.45}               & \multicolumn{1}{c|}{0.36}            & \multicolumn{1}{c|}{0.22}          & \multicolumn{1}{c|}{86\%}              & \multicolumn{1}{c|}{0.86}              & \multicolumn{1}{c|}{0.87}               & \multicolumn{1}{c|}{0.86}            & \multicolumn{1}{c|}{0.83}          \\ \hline
\multicolumn{1}{|c|}{0}                              & \multicolumn{1}{c|}{500}                                & \multicolumn{1}{c|}{500}                           & \multicolumn{1}{c|}{39\%}              & \multicolumn{1}{c|}{0.39}              & \multicolumn{1}{c|}{0.47}               & \multicolumn{1}{c|}{0.39}            & \multicolumn{1}{c|}{0.25}          & \multicolumn{1}{c|}{89\%}              & \multicolumn{1}{c|}{0.89}              & \multicolumn{1}{c|}{0.9}                & \multicolumn{1}{c|}{0.89}            & \multicolumn{1}{c|}{0.86}          \\ \hline
\multicolumn{1}{|c|}{0}                              & \multicolumn{1}{c|}{1000}                               & \multicolumn{1}{c|}{1000}                          & \multicolumn{1}{c|}{45\%}              & \multicolumn{1}{c|}{0.44}              & \multicolumn{1}{c|}{0.55}               & \multicolumn{1}{c|}{0.45}            & \multicolumn{1}{c|}{0.32}          & \multicolumn{1}{c|}{88\%}              & \multicolumn{1}{c|}{0.88}              & \multicolumn{1}{c|}{0.88}               & \multicolumn{1}{c|}{0.88}            & \multicolumn{1}{c|}{0.85}          \\ \hline
\multicolumn{1}{|c|}{0}                              & \multicolumn{1}{c|}{2000}                               & \multicolumn{1}{c|}{2000}                          & \multicolumn{1}{c|}{53\%}              & \multicolumn{1}{c|}{0.5}               & \multicolumn{1}{c|}{0.56}               & \multicolumn{1}{c|}{0.51}            & \multicolumn{1}{c|}{0.39}          & \multicolumn{1}{c|}{\textbf{91\%}}     & \multicolumn{1}{c|}{\textbf{0.91}}     & \multicolumn{1}{c|}{\textbf{0.91}}      & \multicolumn{1}{c|}{\textbf{0.91}}   & \multicolumn{1}{c|}{\textbf{0.89}} \\ \hline
\multicolumn{1}{|c|}{0}                              & \multicolumn{1}{c|}{3000}                               & \multicolumn{1}{c|}{3000}                          & \multicolumn{1}{c|}{51\%}              & \multicolumn{1}{c|}{0.5}               & \multicolumn{1}{c|}{0.59}               & \multicolumn{1}{c|}{0.51}            & \multicolumn{1}{c|}{0.4}           & \multicolumn{1}{c|}{86\%}              & \multicolumn{1}{c|}{0.86}              & \multicolumn{1}{c|}{0.88}               & \multicolumn{1}{c|}{0.86}            & \multicolumn{1}{c|}{0.83}          \\ \hline
\multicolumn{1}{|c|}{0}                              & \multicolumn{1}{c|}{4000}                               & \multicolumn{1}{c|}{4000}                          & \multicolumn{1}{c|}{\textbf{57\%}}     & \multicolumn{1}{c|}{\textbf{0.56}}     & \multicolumn{1}{c|}{\textbf{0.64}}      & \multicolumn{1}{c|}{\textbf{0.57}}   & \multicolumn{1}{c|}{\textbf{0.47}} & \multicolumn{1}{c|}{91\%}              & \multicolumn{1}{c|}{0.9}               & \multicolumn{1}{c|}{0.91}               & \multicolumn{1}{c|}{0.91}            & \multicolumn{1}{c|}{0.88}          \\ \hline
\multicolumn{1}{|c|}{0}                              & \multicolumn{1}{c|}{5000}                               & \multicolumn{1}{c|}{5000}                          & \multicolumn{1}{c|}{55\%}              & \multicolumn{1}{c|}{0.55}              & \multicolumn{1}{c|}{0.63}               & \multicolumn{1}{c|}{0.55}            & \multicolumn{1}{c|}{0.45}          & \multicolumn{1}{c|}{91\%}              & \multicolumn{1}{c|}{0.9}               & \multicolumn{1}{c|}{0.91}               & \multicolumn{1}{c|}{0.91}            & \multicolumn{1}{c|}{0.88}          \\ \hline
\multicolumn{13}{l}{}                                                                           \\ \hline
\multicolumn{1}{|c|}{1000}                           & \multicolumn{1}{c|}{0}                                  & \multicolumn{1}{c|}{0}                             & \multicolumn{1}{c|}{53\%}              & \multicolumn{1}{c|}{0.53}              & \multicolumn{1}{c|}{0.56}               & \multicolumn{1}{c|}{0.53}            & \multicolumn{1}{c|}{0.41}          & \multicolumn{1}{c|}{91\%}              & \multicolumn{1}{c|}{0.91}              & \multicolumn{1}{c|}{0.91}               & \multicolumn{1}{c|}{0.91}            & \multicolumn{1}{c|}{0.89}          \\ \hline
\multicolumn{1}{|c|}{1000}                           & \multicolumn{1}{c|}{500}                                & \multicolumn{1}{c|}{500}                           & \multicolumn{1}{c|}{50\%}              & \multicolumn{1}{c|}{0.51}              & \multicolumn{1}{c|}{0.56}               & \multicolumn{1}{c|}{0.5}             & \multicolumn{1}{c|}{0.38}          & \multicolumn{1}{c|}{93\%}              & \multicolumn{1}{c|}{0.93}              & \multicolumn{1}{c|}{0.93}               & \multicolumn{1}{c|}{0.93}            & \multicolumn{1}{c|}{0.92}          \\ \hline
\multicolumn{1}{|c|}{1000}                           & \multicolumn{1}{c|}{1000}                               & \multicolumn{1}{c|}{1000}                          & \multicolumn{1}{c|}{51\%}              & \multicolumn{1}{c|}{0.52}              & \multicolumn{1}{c|}{0.55}               & \multicolumn{1}{c|}{0.51}            & \multicolumn{1}{c|}{0.4}           & \multicolumn{1}{c|}{91\%}              & \multicolumn{1}{c|}{0.91}              & \multicolumn{1}{c|}{0.91}               & \multicolumn{1}{c|}{0.91}            & \multicolumn{1}{c|}{0.89}          \\ \hline
\multicolumn{1}{|c|}{1000}                           & \multicolumn{1}{c|}{2000}                               & \multicolumn{1}{c|}{2000}                          & \multicolumn{1}{c|}{60\%}              & \multicolumn{1}{c|}{0.6}               & \multicolumn{1}{c|}{0.63}               & \multicolumn{1}{c|}{0.51}            & \multicolumn{1}{c|}{0.5}           & \multicolumn{1}{c|}{92\%}              & \multicolumn{1}{c|}{0.92}              & \multicolumn{1}{c|}{0.92}               & \multicolumn{1}{c|}{0.92}            & \multicolumn{1}{c|}{0.9}           \\ \hline
\multicolumn{1}{|c|}{1000}                           & \multicolumn{1}{c|}{3000}                               & \multicolumn{1}{c|}{3000}                          & \multicolumn{1}{c|}{58\%}              & \multicolumn{1}{c|}{0.58}              & \multicolumn{1}{c|}{0.62}               & \multicolumn{1}{c|}{0.58}            & \multicolumn{1}{c|}{0.48}          & \multicolumn{1}{c|}{91\%}              & \multicolumn{1}{c|}{0.91}              & \multicolumn{1}{c|}{0.92}               & \multicolumn{1}{c|}{0.91}            & \multicolumn{1}{c|}{0.89}          \\ \hline
\multicolumn{1}{|c|}{1000}                           & \multicolumn{1}{c|}{4000}                               & \multicolumn{1}{c|}{4000}                          & \multicolumn{1}{c|}{61\%}              & \multicolumn{1}{c|}{0.61}              & \multicolumn{1}{c|}{0.64}               & \multicolumn{1}{c|}{0.61}            & \multicolumn{1}{c|}{0.51}          & \multicolumn{1}{c|}{\textbf{93\%}}     & \multicolumn{1}{c|}{\textbf{0.93}}     & \multicolumn{1}{c|}{\textbf{0.94}}      & \multicolumn{1}{c|}{\textbf{0.93}}   & \multicolumn{1}{c|}{\textbf{0.92}} \\ \hline
\multicolumn{1}{|c|}{1000}                           & \multicolumn{1}{c|}{5000}                               & \multicolumn{1}{c|}{5000}                          & \multicolumn{1}{c|}{\textbf{62\%}}     & \multicolumn{1}{c|}{\textbf{0.62}}     & \multicolumn{1}{c|}{\textbf{0.68}}      & \multicolumn{1}{c|}{\textbf{0.62}}   & \multicolumn{1}{c|}{\textbf{0.53}} & \multicolumn{1}{c|}{93\%}              & \multicolumn{1}{c|}{0.93}              & \multicolumn{1}{c|}{0.93}               & \multicolumn{1}{c|}{0.93}            & \multicolumn{1}{c|}{0.91}          \\ \hline
\end{tabular}
\end{adjustbox}
\caption{Classification results of the ViT\_b\_16 model on the test dataset (comprised of real images), trained using various combinations of real and synthetic images.}
\label{tab:vit}
\end{table}

\begin{table}[ht]
\centering
\begin{tabular}{|c|c|c|c|c|c|c|c|}
\hline
\textbf{Real} & \textbf{StyleGAN} & \textbf{LDM} & \textbf{Accuracy} & \textbf{F1 Score} & \textbf{Precision} & \textbf{Recall} & \textbf{MCC} \\
\hline
0 & 0 & 250  & 47\% & 0.52 & 0.60 & 0.47 & 0.40 \\\hline
0 & 0 & 500  & 67\% & 0.69 & 0.74 & 0.66 & 0.61 \\\hline
0 & 0 & 1000 & 87\% & 0.87 & 0.88 & 0.87 & 0.84 \\\hline
0 & 0 & 2000 & 90\% & 0.90 & 0.90 & 0.90 & 0.88 \\\hline
0 & 0 & 3000 & 87\% & 0.87 & 0.89 & 0.87 & 0.84 \\\hline
0 & 0 & 4000 & 89\% & 0.89 & 0.89 & 0.90 & 0.85 \\\hline
0 & 0 & 5000 & 89\% & 0.89 & 0.89 & 0.90 & 0.86 \\\hline
\end{tabular}
\caption{Classification results of the VGG16 model on the test dataset (consisting of real images), trained using various amounts of LDM-generated images.}
\end{table}

\begin{table}[ht]
\centering
\begin{tabular}{|c|c|c|c|c|c|c|c|}
\hline
\textbf{Real} & \textbf{StyleGAN} & \textbf{LDM} & \textbf{Accuracy} & \textbf{F1 Score} & \textbf{Precision} & \textbf{Recall} & \textbf{MCC} \\
\hline
0 & 250  & 0 & 38\% & 0.27 & 0.22 & 0.38 & 0.24 \\\hline
0 & 500  & 0 & 47\% & 0.35 & 0.28 & 0.47 & 0.37 \\\hline
0 & 1000 & 0 & 48\% & 0.41 & 0.36 & 0.48 & 0.38 \\\hline
0 & 2000 & 0 & 53\% & 0.52 & 0.52 & 0.53 & 0.45 \\\hline
0 & 3000 & 0 & 51\% & 0.38 & 0.31 & 0.51 & 0.42 \\\hline
0 & 4000 & 0 & 53\% & 0.53 & 0.53 & 0.53 & 0.45 \\\hline
0 & 5000 & 0 & 53\% & 0.57 & 0.63 & 0.53 & 0.46 \\\hline
\end{tabular}
\caption{Classification results of the VGG16 model on the test dataset (consisting of real images), trained using various amounts of StyleGAN-generated images.}
\end{table}
\renewcommand{\thefigure}{S\arabic{figure}}
\begin{figure}[htb!]
\centering
\includegraphics[width=\linewidth]{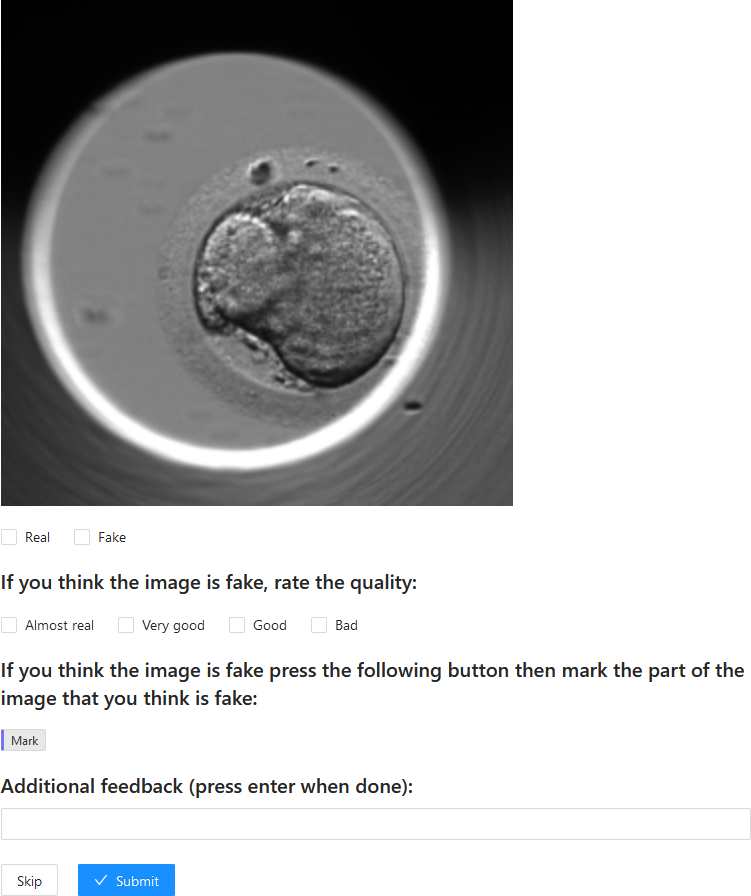}
\caption{The user interface of the web application as presented to subject matter experts.}
\label{fig:webapp}
\end{figure}